%% file: main_vldb_icdt.tex
\documentclass[runningheads]{llncs}

\usepackage[T1]{fontenc} 

\usepackage{graphicx}

\usepackage[appendix=inline]{apxproof} 
\usepackage{fancyvrb}
\usepackage{paralist}

\theoremstyle{remark} 

\theoremstyle{definition}
\newtheorem{notation}{Notation}

\usepackage[inline]{enumitem}

\usepackage{amsmath,amsthm,amsfonts}
\usepackage{stmaryrd}
\usepackage{url}

\usepackage{mathtools}
\DeclarePairedDelimiter{\ceil}{\lceil}{\rceil}

\usepackage{color,colortbl}
\definecolor{Gray}{gray}{0.9}
\definecolor{LightCyan}{rgb}{0.88,1,1}

\usepackage{booktabs}

\usepackage{xcolor}

\newtheoremrep{lemmax}[lemma]{Lemma}
\newtheoremrep{theox}[theorem]{Theorem}
\newtheoremrep{corx}[corollary]{Corollary}

 \input{macros.tex}

 \newcommand{\xfp}{fix-point}

\title{Negation-Closure for JSON Schema}

\author{Mohamed-Amine Baazizi\inst{1}
\and Dario Colazzo\inst{2}
\and Giorgio Ghelli\inst{3}
\and Carlo Sartiani\inst{4}
\and Stefanie Scherzinger \inst{5}
}
\authorrunning{Baazizi et al.}
\institute{Universit\'e Paris-Dauphine, PSL Research University, France\\
\email{\{dario.colazzo\}@dauphine.fr}
\and
Sorbonne Universit\'e, LIP6 UMR 7606, France
\email{baazizi@ia.lip6.fr}
\and
Universit\'e Paris-Dauphine, PSL Research University, France\\
\email{\{dario.colazzo\}@dauphine.fr}
\and
Dipartimento di Informatica, Universit\`a di Pisa, Italy
\email{ghelli@di.unipi.it}
\and
DIMIE, Universit\`a della Basilicata, Italy
\email{carlo.sartiani@unibas.it}
\and
University of Passau, Passau, Germany \\
\email{\{stefanie.scherzinger\}@uni-passau.de}
}
\begin{document}

\maketitle

\begin{abstract}

{\jsonsch} is an evolving standard for describing families of {\json}  documents. It is a logical language, based on a set of \emph{assertions} that describe features of the {\json} value under analysis and on logical or structural combinators for these assertions, including a negation operator. 
Most logical languages with negation enjoy \emph{negation closure}, that is, for every operator they have a negation dual
that expresses its negation.
We show that this is not the case for {\jsonsch}, we study how that changed with the latest versions of the Draft,
and we discuss how the language may be enriched accordingly.
In the process, we define an algebraic reformulation of JSON Schema, which we successfully employed in a prototype system for generating schema witnesses.
\end{abstract}



\section{Introduction}\label{sec:intro}
\input{introNew.tex}


\section{An algebraic version of {\jsonsch}}\label{sec:full}

\input{jsonschemaalgebra.tex}

\section{Negation closure}\label{sec:notelimination}
\input{negationclosure.tex}

\begin{example}\label{ex:notelim}
\input{examples}

\end{example}

\hide{
\section{Core algebra}\label{sec:core}
\input{core.tex}

}

\section{Going towards {\VerNine}}\label{sec:mincontains}
\input{towardsnewver.tex}

\section{Experiments}\label{sec:experiments}
\input{experiments.tex}

\section{Related work}
\label{sec:related}
\input{relatedwork.tex}

\section{Conclusions}

We have shown that {\jsonsch} is ``almost'' negation-closed 
and we have provided an exact characterization of the schemas that cannot be expressed without negation. 
We have studied the impact of the new operators $\xminC$ and $\xmaxC$ introduced in {\VerNine},
and we have shown that they make the array construct negation-closed, at the price of a non-trivial encoding
and exponential-size explosion.
We have introduced an algebraic rendition of {\jsonsch} syntax that is amenable for automated manipulation.

We contributed a not-elimination algorithm that we are currently using as a building block for our algorithm
for witness generation, satisfiability checking, and inclusion verification~\cite{DBLP:conf/edbt/AttoucheBCFGLSS21}.
As shown in Example~\ref{ex:notelim}, not-elimination can also be useful to improve the readability of some {\jsonsch}
documents.
Our not-elimination algorithm is the first that deals with negative recursive variables, and it uses an original, and very simple, 
technique to do so.

To check the completeness of our approach, we have implemented the translation from JSON Schema 
to the algebra and back, as well as a version of the not-elimination algorithm, and we tested them on our 91K collected schemas 
(after data cleaning and 
duplicate-elimination). The experiments
confirmed that we deal with every aspect of the language, and that the not-elimination result has a size that
grows linearly with the input size.
The implementations are available,\footnote{\url{https://jsonschematool.ew.r.appspot.com/}} along with a very preliminary version of witness generation, 
and our schemas~\cite{schema_corpus}.



\bibliographystyle{splncs04}
\bibliography{references}

\end{document}

%% file: macros.tex
\usepackage{tikz}
\usepackage{ulem}
\newcommand*\circled[1]{\tikz[baseline=(char.base)]{
            \node[shape=circle,draw,inner sep=0.5pt] (char) {#1};}}

  {\small \em \tt ccc\begin{table} aaa bbb}%
  {\end{table} ddd}
  
\newcommand{\Tr}[1]{\BeginTr{#1}\EndTr}  
\newcommand{\BeginTr}{\ensuremath{\langle}}
\newcommand{\EndTr}{\ensuremath{\rangle}}

\newcommand{\json}{JSON}
\newcommand{\kw}[1]{\textbf{#1}}
\renewcommand{\kw}[1]{\ensuremath{\mathtt{#1}}}
\newcommand{\qkw}[1]{\ensuremath{\mathtt{\QQ{#1}\QQ}}}
\newcommand{\key}[1]{\ensuremath{\mathit{#1}}}

\newcommand{\akey}[1]{\ensuremath{\mathsf{#1}}}
\newcommand{\qakey}[1]{\ensuremath{\mathsf{\QQ{#1}\QQ}}}

\newcommand{\xnot}{\kw{not}}

\newcommand{\xtrue}{\kw{true}}

\newcommand{\xfalse}{\kw{false}}

\newcommand{\xnull}{\kw{null}}

\newcommand{\xone}{\kw{oneOf}}

\newcommand{\xany}{\kw{anyOf}}

\newcommand{\xall}{\kw{allOf}}

\newcommand{\xmin}{\kw{minimum}}

\newcommand{\xmax}{\kw{maximum}}

\newcommand{\xreq}{\kw{required}}

\newcommand{\xtype}{\kw{type}}

\newcommand{\xexmin}{\kw{exclusiveMinimum}}

\newcommand{\xexmax}{\kw{exclusiveMaximum}}

\newcommand{\xprops}{\kw{properties}}
\newcommand{\qprops}{\qkw{properties}}
\newcommand{\xpropN}{\kw{propertyNames}}

\newcommand{\xpattProps}{\kw{patternProperties}}

\newcommand{\xminP}{\kw{minProperties}}

\newcommand{\xmaxP}{\kw{maxProperties}}

\newcommand{\xthen}{\kw{then}}

\newcommand{\xif}{\kw{if}}

\newcommand{\xelse}{\kw{else}}

\newcommand{\xite}{\xif-\xthen-\xelse}

\newcommand{\xaddProps}{\kw{additionalProperties}}
\newcommand{\qaddProps}{\qkw{additionalProperties}}

\newcommand{\xaddIts}{\kw{additionalItems}}
\newcommand{\qaddIts}{\qkw{additionalItems}}

\newcommand{\xmof}{\kw{multipleOf}}

\newcommand{\xmaxL}{\kw{maxLength}}

\newcommand{\xminL}{\kw{minLength}}

\newcommand{\xpatt}{\kw{pattern}}

\newcommand{\xuniqIt}{\kw{uniqueItems}}

\newcommand{\xcont}{\kw{contains}}

\newcommand{\xminC}{\kw{minContains}}

\newcommand{\xmaxC}{\kw{maxContains}}

\newcommand{\xminIt}{\kw{minItems}}

\newcommand{\xmaxIt}{\kw{maxItems}}

\newcommand{\xit}{\kw{items}}
\newcommand{\qit}{\qkw{items}}
\newcommand{\xdepS}{\kw{dependentSchemas}}

\newcommand{\xdeps}{\kw{dependencies}}

\newcommand{\xdepR}{\kw{dependentRequired}}

\newcommand{\xenum}{\kw{enum}}

\newcommand{\xconst}{\kw{const}}

\newcommand{\xdref}{\kw{\$ref}}
\newcommand{\qdref}{\qkw{\$ref}}

\newcommand{\xdefault}{\kw{default}}

\newcommand{\xdefs}{\kw{definitions}}

\newcommand{\xddefs}{\kw{\$defs}}

\newcommand{\xobject}{\akey{object}}
\newcommand{\qobject}{\qakey{object}}

\newcommand{\xid}{\kw{id}}

\newcommand{\xdid}{\kw{\$id}}

\newcommand{\xdschema}{\kw{\$schema}}

\newcommand{\xda}{\kw{\$anchor}}

\newcommand{\ADefKey}{\akey{root}}

\newcommand{\DefWithKey}[3]{{#1}{#2}:{#3}}
\newcommand{\Def}[2]{\DefWithKey{\DefKey\ }{#1}{#2}}
\renewcommand{\Def}[2]{\DefWithKey{}{#1}{#2}}

\newcommand{\Defs}{\akey{defs}}
\newcommand{\ADef}[2]{\DefWithKey{\ADefKey\ }{#1}{#2}}
\renewcommand{\ADef}[2]{\DefWithKey{\ADefKey\ }{#1}{#2}}

\newcommand{\RRef}[1]{\key{#1}}

\newcommand{\defAss}{definitional assertion}

\newcommand{\gcomment}[1]{}
\newcommand{\oldversion}[1]{}
\newcommand{\hideforspace}[1]{}
\newcommand{\hide}[1]{}
\newcommand{\code}[1]{}

\newcommand{\Iff}{\Leftrightarrow}
\newcommand{\Implies}{\Rightarrow}
\newcommand{\RevImplies}{\Leftarrow}
\newcommand{\Else}{\ |\ }
\newcommand{\Or}{\vee}
\newcommand{\BigOr}{\bigvee}

\renewcommand{\And}{\wedge}
\newcommand{\BigAnd}{\bigwedge}
\newcommand{\XOr}{\oplus}
\renewcommand{\XOr}{\circled{\textup{\small 1}}}

\newcommand{\Not}{\neg}
\newcommand{\True}{{\bf t}}
\newcommand{\False}{{\bf f}}
\newcommand{\Type}{\akey{type}}
\newcommand{\Num}{\akey{Num}}
\newcommand{\Str}{\akey{Str}}
\newcommand{\Int}{\akey{Int}}
\newcommand{\Arr}{\akey{Arr}}
\newcommand{\Obj}{\akey{Obj}}
\newcommand{\Bool}{\akey{Bool}}
\newcommand{\Null}{\akey{Null}}
\newcommand{\TT}[1]{\Type({#1})}
\newcommand{\TNum}{\TT{\Num}}
\newcommand{\TStr}{\TT{\Str}}

\newcommand{\TInt}{\TT{\Int}}
\newcommand{\TArr}{\TT{\Arr}}
\newcommand{\TObj}{\TT{\Obj}}
\newcommand{\TBool}{\TT{\Bool}}

\newcommand{\Mof}{\akey{mulOf}}
\newcommand{\NotMof}{\akey{notMulOf}}
\newcommand{\Pat}{\akey{pattern}}
\newcommand{\NotPat}{\akey{notPattern}}
\newcommand{\Uni}{\akey{uniqueItems}}
\newcommand{\NotUni}{\akey{repeatedItems}}
\newcommand{\CProp}[2]{\akey{prop}(\key{#1}:{#2})}

\newcommand{\Props}{\akey{props}}
\newcommand{\CIte}{\akey{ite}}
\newcommand{\CItem}[2]{\akey{item}(#1:#2)}
\newcommand{\CItems}[2]{\akey{items}(#1\uparrow:#2)}
\newcommand{\IteK}{\akey{items}}
\newcommand{\Ite}[2]{\IteK(#1;#2)}
\newcommand{\itdots}{,\ldots,}

\newcommand{\Bet}{\akey{betw}}
\newcommand{\XBet}{\akey{xBetw}}
\newcommand{\Len}{\akey{len}}
\newcommand{\Ex}{\#}
\newcommand{\Cont}[1]{\Ex_1^{\Inf}{#1}}
\renewcommand{\Cont}[1]{\ContK({#1})}
\newcommand{\ContK}{\akey{contains}}
\newcommand{\ContAfterK}{\akey{contAfter}}
\newcommand{\ContAfter}[2]{\ContAfterK({#1}:{#2})}

\newcommand{\Pro}{\akey{pro}}
\newcommand{\Nam}{\akey{pNames}}
\newcommand{\ExNam}{\akey{exPName}}
\newcommand{\keykey}[1]{\key{\underline{#1}}}
\newcommand{\Con}{\akey{const}}
\newcommand{\Enu}{\akey{enum}}

\newcommand{\IBT}{\akey{isBoolValue}}
\renewcommand{\IBT}{\akey{ifBoolThen}}
\newcommand{\Inf}{\infty}

\newcommand{\PReq}{\akey{pattReq}}
\newcommand{\APReq}{\akey{addPattReq}}

\newcommand{\Req}{\akey{req}}
\newcommand{\AddP}{\akey{addp}}

\newcommand{\VerSix}{Draft-06}

\newcommand{\VerEight}{Draft 2019-09}
\newcommand{\VerNine}{Draft 2019-09}
\newcommand{\JS}{JSON Schema}
\newcommand{\Nat}{\mathbb{N}}

\newcommand{\POfS}{\key{PattOfS}}
\newcommand{\TrueP}{\True^{\bullet}}
\newcommand{\FalseP}{\False^{\bullet}}
\newcommand{\AndP}{\And^{\bullet}}

\newcommand{\NotP}{\Not^{\bullet}}

\newcommand{\To}{\rightarrow}

\newcommand{\custcom}[2]{\marginpar{\tiny #1: {#2}}}
\renewcommand{\custcom}[2]{}

\newcommand{\GG}[1]{\custcom{giorgio}{#1}}
\newcommand{\DC}[1]{\custcom{dario}{#1}}

\newcommand{\M}{\ |\ }

\newcommand{\Sum}{\key{sum}}
\newcommand{\NotIf}{\key{NotIf}}
\newcommand{\SIf}{\key{Iff}}

\newlength{\NL}
\newlength{\SaveNL}

\newcommand{\NotVarFun}{\key{co}}
\newcommand{\NotVar}[1]{\NotVarFun(\key{#1})}

\newcommand{\NN}{\ensuremath{\ \hat{}\ }}
\newcommand{\QQ}{\textnormal{\textquotedbl}}
\newcommand{\Set}[1]{\{\,{#1}\,\}}
\newcommand{\SetTo}[1]{\{1..{#1}\}}
\newcommand{\SetFromTo}[2]{\{{#1}..{#2}\}}

\newcommand{\NR}[1]{\RRef{\NotVar{#1}}}

\newcommand{\JU}{\mathcal{J}}

\newcommand{\nule}{\nullv}

\newcommand{\nullv}{\ensuremath{\text{null}}}
\newcommand{\str}{s}

\newcommand{\J}{{J}}

\newcommand{\truev}{\ensuremath{\text{true}}}
\newcommand{\falsev}{\ensuremath{\text{false}}}
\newcommand{\numm}{m}

\newcommand{\semt}{\ensuremath{\mathit{JVal}}}
\newcommand{\semcapar}[3]{[\![ #1 ]\!]_{#2}^{#3}}
\newcommand{\semcai}[2]{\semcapar{#1}{#2}{i}}
\newcommand{\semca}[2]{[\![ #1 ]\!]_{#2}}
\newcommand{\E}{E}
\newcommand{\e}{\epsilon}

\newcommand{\setst}[2]{\{ #1 \ | \ #2\}}
\newcommand{\rlan}[1]{L(#1)}
\newcommand{\noa}[1]{|#1|}

\newcommand{\Inv}[1]{#1^{-1}}

\newcommand{\TGIN}[1]{\akey{#1}}
\newcommand{\TG}[1]{\{\TGIN{#1}\}}

\newcommand{\TSt}{\TG{\SSt}}

\newcommand{\SSt}{{Str}}

\newcommand{\jsonsch}{JSON Schema} 

\newcommand{\RegExp}{\mathit{RegExp}}
\newcommand{\MNum}{\mathbb{R}}
\newcommand{\JSet}{\mathbb{J}}

\newcommand{\xast}[1]{\QQ #1\QQ}


%% file: introNew.tex

{\json} is a simple data  language whose terms represent trees constituted by nested records and arrays, with atomic values at the leaves, and is now widely used for data exchange on the Web.
{\jsonsch}~\cite{jsonschema} is an ever evolving specification for describing families of {\json} terms,
and heavily used for specifying web applications and REST services~\cite{DBLP:conf/issta/HabibSHP21}, 
compatibility of operators in data science pipelines~\cite{lale2020}, as well as  schemas in NoSQL systems 
(e.g., MongoDB~\cite{mongodbschema}).

{\jsonsch} is a logical language based on a set of \emph{assertions} that describe features of {\json} 
values, and on boolean and structural combinators for these assertions, including negation and recursion. 
The expressive power of this complex language, and the complexity of validation and satisfiability have recently been studied:
Pezoa et al.~\cite{DBLP:conf/www/PezoaRSUV16} relied on tree automata and MSO to study the expressive power, while Bourhis et al.~\cite{DBLP:conf/pods/BourhisRSV17} mapped 
{\jsonsch} onto an equivalent modal logic, called JSL,  
to investigate the complexity of validation and satisfiability, and proved that satisfiability is in 2EXPTIME in the general case.

In this context, we are working on tools for the simplification and manipulation of {\jsonsch},
and we have studied the problem of \emph{negation-closure}, that is the property
that every negated assertion can be rewritten into a negation-free one.
Most logical languages with negation enjoy \emph{negation-closure} because, for every operator, they have a 
``dual'' that allows negation to be pushed to the leaves of any logical formula, as happens for the pairs
\emph{and-or} and \emph{forall-exists} in first order logics, and for the modal-logic pair
$\Diamond$-$\Box$ used in JSL to encode {\jsonsch}.
Negation-closure is an important design principle for a logical system, since it ensures that, for every algebraic 
property that involves an operator, a ``symmetric'' algebraic property holds for its dual operator.
This facilitates both reasoning and
automatic manipulation.

In this paper, we prove that {\jsonsch}, despite having the same expressive power as JSL, does not enjoy 
negation-closure,  by  showing  that  both  objects and arrays are described by pairs 
of operators that are ``almost'' able to describe the negation of the other one, but not ``exactly''.
We mostly focus here on {\VerSix}~\cite{Draft06} of the JSON Schema language. Indeed, we have built a repository of JSON Schema documents
(available at~\cite{schema_corpus}), by harvesting 
from GitHub all JSON documents with a {\xdschema} attribute, retrieving over 91K documents; we found that almost all of them adhere to {\VerSix}  or to a preceding one.

We show that, in the most common use-cases, negation can indeed be pushed through these operators,
and we exactly characterize the specific cases when this is not possible.
Moreover, we show that  {\VerEight} introduces another twist:
a new version of the \xcont\ 
operator allows not-elimination for array schemas in the general case, but only
through a complex encoding.

In addition,  based on these results, we define and present here a negation-closed  extension of {\jsonsch}, with the same expressive
power as the original language, but where all operators have a negation dual, together  with a simple and complete not-elimination algorithm for the extended language.

Our recent witness generation tool for JSON Schema~\cite{DBLP:conf/edbt/AttoucheBCFGLSS21} 
is capable of generating a valid instance, given a satisfiable schema. That approach is able to deal with almost the totality of JSON Schema (the only mechanism we rule out is {\Uni}), and to this end it proceeds in an inductive fashion: first it generates witnesses for subexpressions and then these are used for witness generation of surrounding expressions. In this context, one problem to solve is negation: there is no way to generate a witness for $\Not S$ starting from a witness 
for an arbitrarily complex schema $S$.
To address this issue, our witness generation  approach heavily relies on the not-elimination algorithm we present here,  and we believe that the present study sheds light on aspects of JSON Schema that have not been studied in previous works \cite{DBLP:conf/www/PezoaRSUV16,DBLP:conf/pods/BourhisRSV17}, 
and that may be useful for the development of tools manipulating JSON Schemas.

\DC{in previous reviews, reviewers said they had doubts about the usefulness of not-elimination, so I added explanations here hoping that this time is much more clear. }

\smallskip
\noindent
\textbf{Main Contributions:} 

\begin{compactitem}
\item [(i)]
We study the problem of negation-closure of {\jsonsch}, that is, which operators are endowed with a negation
dual. 
We show that {\jsonsch} structural operators are \emph{not} negation-closed, and we characterize
the schemas whose negation cannot be expressed without negation.
In the process, we present a reformulation of {\jsonsch} that is \emph{algebraic}, that is, where a subschema can be freely substituted by an equivalent one.
%

\item [(ii)]
We then extend the algebraic version of {\jsonsch} to make it negation-closed, and we 
define a not-elimination algorithm for this closed algebra.
To our knowledge, this is the first algorithm for not-elimination that also deals with negated recursive
variables.


\item[(iii)]
We extend our study to the latest drafts of {\jsonsch},
which introduced operators that impact negation closure.

\end{compactitem}


\smallskip
\noindent
\textbf{Paper Outline:} 
In Section~\ref{sec:full} we present an algebraic version {\jsonsch},
and its formal semantics.
Section~\ref{sec:notelimination} studies \emph{negation closure}. 
Section~\ref{sec:mincontains} introduces the latest {\jsonsch} draft.
Section~\ref{sec:experiments} presents our experiments. 
We conclude after a review of related work.

%% file: jsonschemaalgebra.tex

In this work we rely on an algebraic reformulation of JSON Schema, that has a direct correspondence with JSON Schema, but that is more amenable for formal development of schema  manipulation algorithms.  Also, and  importantly, our algebra enjoys  substitutability, in the sense that the substitution of every subterm in an algebraic term with a semantically equivalent  subterm preserves the semantics of the entire term, independently of 
the context around the subterm. Actually, current formulations of JSON Schema do not enjoy this property, while it is fundamental in the design of schema manipulation algorithms.  

In the remaining part of this section we first define the JSON data model and then our algebraic definitions of JSON Schema. 


\subsection{The data model}\label{subsec:dm}

{\json} values are either basic values, objects, or arrays.
Basic values~$B$ include  the \nule\ value, booleans, numbers $\numm$, and strings $\str$.
Objects~$O$ represent sets of \emph{members}, each member (or \emph{field}) being a name-value pair $(k,\J)$, and arrays $A$ represent sequences of values with positional access.
%
We will only consider here objects without repeated names. 
Objects and arrays may be empty.

In \json\ syntax, a name is itself a string, and hence it is surrounded by quotes.
%
Below, we specify the data model syntax, where $n$ is a natural number with $n\geq 0$, and $k_i\in\Str$ for $i = 1, \ldots, n$.
\[
\begin{array}{llrllllllll}
\J ::= & B \mid O \mid A   & &\text{\bf {\json} expressions}  \\
B ::=	 & \xnull \mid \xtrue \mid \xfalse \mid \numm \mid \str  & \numm\in\Num, \str\in\Str  &  \text{\bf Basic values} \\ 
O ::= & \kw{ \{} k_1:\J_1,\ldots,k_n:\J_n \kw{\}}\   &  \ n\geq 0, \  i\neq j \Rightarrow k_i\neq k_j & \text{\bf Objects}  \\
A ::= &  \kw{[} \J_1, \ldots, \J_n\kw{]} \  & n\geq 0  &  \text{\bf Arrays} 
\end{array}
\]

\subsection{The algebra}
We now introduce 
our algebraic presentation of {\jsonsch}. The syntax of the algebra is shown below.
%
In the grammar $n$ is always a natural number starting from $0$, so that we use here indexes going from $1$ to $n+1$ when
at least one element is required. 
In $\Req(k_1,\ldots,k_n)$, each $k_i$ is a string. The other metavariables are listed in the first line of the grammar,
where $\MNum$ indicates ``any real number'', $\MNum_{>0}$ denotes positive real numbers, $\Nat$ are the naturals zero included, and the sets $\MNum^{-\Inf}$,
$\MNum^{\Inf}$, and $\Nat^{\Inf}$ stand for the base set enriched with the extra symbols $- \Inf$ or~$\Inf$. $\JSet$ is the set of all
{\json} terms.
Apart from the syntax, the algebra reflects all operators of {\jsonsch}, {\VerSix}.
\setlength{\SaveNL}{\NL}
\setlength{\NL}{0.7ex}
\[
\small
\begin{array}{llll}
\multicolumn{4}{l}{r \in \RegExp, 
m\in\MNum^{-\Inf}, M\in\MNum^{\Inf},
 l\in\Nat, j \in \Nat^{\Inf}, q \in \MNum_{>0},
 J \in \JSet }\\[\NL]

T & ::= & \Arr \M \Obj \M \Null \M \Bool \M \Str  \M \Num  \\[\NL] 
S & ::= &\ \Type(T_1,\ldots,T_{n+1}) \M  \Con(J) \M \Enu(J_1,\ldots,J_{n+1}) \M \Len_{l}^{j} \M \Bet_{m}^{M}  \\[\NL] 
& & \M \XBet_{m}^{M}  \M \Mof(q)  \\[\NL]
&&  \M \Pat(r) \M \Props(r_1 : S_1,\ldots,r_n : S_n; S)   \M \Pro_{l}^{j} \M \Req(k_1,\ldots,k_n) \\[\NL]
& &  \M \Nam(S) \\[\NL]
&&  \M \Ite{S_1 \itdots  S_n}{S_{n+1}} 
     \M \Cont{S}   \M \CIte_l^j  \M \Uni  
       \\[\NL]
&&  \M \RRef{x}  \M \True  \M  \False \M \Not S \M  S_1 \Or S_2   \M \XOr(S_1,\ldots,S_{n+1})  \\[\NL]
& &  \M S_1 \Implies S_2 \M  ( S_1 \Implies S_2 \ | \ S_3 )
 \M S_1 \And S_2 \M \{ S_1, \ldots, S_n \} \\[\NL] 
\E & ::= & \Def{x_1}{S_1} , \ldots, \Def{x_{n}}{S_{n}} \\[\NL]
D & ::= & S\ \Defs\ (\E) \\[\NL]
\end{array}
\]

\setlength{\NL}{\SaveNL}


Each schema expresses properties of an \emph{instance} which is a {\json}  value;  
the semantics of a schema $S$ with respect to an environment~$\E$, hence, is the set   $\semca{S}{\E}$ of {\json}  instances 
that \emph{satisfy} that schema, as specified
in Figure~\ref{fig:fullsem-ws}. 
The environment ${\E}$ is a set of pairs $(\Def{x}{S})$,
that are introduced by the operator $D = S\ \Defs(\Def{x_1}{S_1} , \ldots, \Def{x_{n}}{S_{n}}) $
and are  used to interpret variables~$x_i$, as discussed below.

The schema  $\Type(T_1,\ldots,T_{n+1})$ is satisfied by any
instance belonging to one of the listed predefined JSON types. 

The algebra includes boolean operators ($\Not$, $\And$, $\Or$, $\Implies$, $\Implies \ |$, $\{ S_1, \ldots, S_n \}$, and $\XOr$) as well as \emph{typed} operators (the remaining ones, each one related to one type $T$, with the exception of $\Null$)), whose semantics is described below.

$\Con(J)$ is only satisfied by the instance $J$, and $\Enu(J_1,\ldots,J_{n+1})$ is the same as $\Con(J_1)\Or\ldots\Or\Con(J_{n+1})$.

 $\Pat(r)$ means: \emph{if} the instance is a string, \emph{then} it matches~$r$.
This conditional semantics
is a central feature of {\jsonsch}: all operators related to one specific type, that is, all operators
from $ \Len_{l}^{j}$ to $\Uni$, have an if-then-else semantics, with the if part 
always: ``if the instance belongs to the type associated with this assertion'',
so that they discriminate inside their type but accept any instance of any other type.
$\Len_{l}^{j}$ means: if the instance is a 
string, then
its length is included between $l$ and $j$. 
$\Bet_{m}^{M}$ means: if the
instance is a number, then it is included between $m$ and $M$, extremes included. $\XBet_{m}^{M}$ is the same with extremes excluded.
$\Mof(q)$ means: if the instance $\J$ is a number, then $\J=i*q$, for some integer $i$.



An instance $J$ satisfies  the assertion $\Props(r_1 : S_1,\ldots,r_n : S_n; S)$ iff the following holds: if the instance $J$ is an object, then for each pair $k:J'$ appearing at the top level of $J$, then, for every $r_i : S_i $ such that $k$ matches $r_i$, then  $J'$ satisfies~$S_i$, and, when $k$ does not match any pattern in $r_1,\ldots,r_n$,  then $J'$ satisfies $S$
(hence, it combines the three {\jsonsch} operators \xprops, \xpattProps, and \xaddProps).

$\Pro_{l}^{j}$ means: if the instance is an object, it has at least $l$  and at most $j$ properties.
Assertion $\Req(k_1,\ldots,k_n)$ means: if the instance is an object, then, for each $k_i$, one of the names of the instance
is equal to $k_i$.
The assertion $\Nam(S)$ means that, if the instance is an object, then every member name of that object satisfies~$S$.

\hide{
The assertion $\PReq(r_1 : S_1,\ldots,r_n : S_n)$ means: if the instance is an object, then, for each $r_i$, at least one of the names of the instance matches $r_i$ and its value satisfies~$S_i$.
The assertion $\APReq((r_1,\ldots,r_n) : S)$ means: if the instance is an object, then it contains at least
one name that does not match any $r_i$ and whose value satisfies $S$.

The assertion $\ExNam(S)$ means that, if the instance is an object, then at least one of its names 
satisfies $S$.

\begin{remark}
Names in $\Req(k_1,\ldots,k_n)$ have no associated type, while in $\PReq(r_1 : S_1,\ldots,r_n : S_n)$
a type must be associated to each pattern.
This happens since, when a pattern $r$ only matches one name, $\PReq(r : S)$ is the same as 
$\PReq(r : \True) \And \CProp{r}{S}$ (where $\True$ is satisfied by any instance),
hence the type is not needed.
In the general case, however, $\PReq(r : \True)\And \CProp{r}{S}$ requires that one
name that matches $r$ exists and that all such names satisfy $S$, while $\PReq(r : S)$
is weaker, since it only requires that \emph{one} of those names satisfies $S$.
To sum up, in order to express negation of $\Props(r:S;\True)$, the assertion $\PReq(r : \True)\And r: \Not S$ would be too
strong, we need $\PReq(r : \Not S)$.
\end{remark}
}

An instance $J$ satisfies $\Ite{S_1 \itdots  S_n}{S_{n+1}}$ iff the following holds: if $J$ is an array, then each of its elements at position $i\leq n $ satisfies $S_i$, while further elements satisfy $S_{n+1}$. Note that no constraint is posed over the length of $J$: if it is strictly shorter than~$n$, or empty, that is not a problem
(this operator combines the two {\jsonsch} operators \xit\ and \xaddIts).
To constrain the array length we have $\CIte_l^j 
$, satisfied by $J$ when: if $J$ is an array, its length is between $l$ and $j$.  
Assertion $\Cont{S}$ means: if the instance is an array, then it contains at least one element that satisfies $S$.
The assertion $\Uni$ means that, if the instance is an array, then all of its items are pairwise different.

The boolean operators $\True$, $\False$, $\Not$, $\Or$, $\Implies$, $\And$ combine the results of their operands in the standard way,
while $( S_1 \Implies S_2 \ | \ S_3 )$ stands for $(S_1 \And S_2) \Or ((\Not S_1) \And S_3)$,
and $\XOr(S_1,\ldots,S_{n+1})$ is satisfied iff exactly one of the arguments holds.  $\{ S_1, \ldots, S_n \}$ 
is the same as $S_1 \And \ldots \And S_n$. 


%


An instance $J$ satisfies a \emph{schema document} $D = S\ \Defs(\E)$
iff it satisfies $S$ in the environment $\E = \Def{x_1}{S_1} , \ldots, \Def{x_{n}}{S_{n}}$,
which means that, when a variable $x_i$ is met while checking whether $J$ satisfies $S$,
$x_i$ is substituted by $\E(x_i)$, that is, by $S_i$.
This is not an inductive definition, since~$S_i$ is generally bigger than $x_i$, and results in a cyclic definition when
we have environments such as $\Defs(\Def{x}{x})$ or $\Defs(\Def{x}{\Not x})$.
The JSON Schema standard rules out such environments by specifying that checking whether $J$ satisfies~$S$ by
expanding variables must never result in an infinite loop.
This can be ensured by imposing a guardedness condition on $\E$, as follows. 

Let us say that $x_i$ 
\emph{unguardedly depends} on $x_j$ if the definition of $x_i$ contains one occurrence of $x_j$ that is not
in the scope of any typed operator
(otherwise, we say that the occurrence is \emph{guarded} by a typed operator): for instance, in 
$\Defs(\ \Def{x}{(\Ite{y}{\RRef{w}} \And  \RRef{z} })\ )$, $x$ unguardedly depends on $z$, while $y$ and $w$ are guarded by $\Ite{}{}$.
Recursion is \emph{guarded} if the \emph{unguardedly depends} relation is acyclic: no pair $(x,x)$ belongs
to its transitive closure. 
Informally, guarded recursion requires that any cyclic dependency must traverse a \emph{typed} operator,
which ensures that, when a variable is unfolded for the second time, the resulting schema is applied to an instance that is strictly
smaller than the one analyzed at the moment of the previous unfolding.
This notion was introduced as \emph{well-formedness} in related work~\cite{DBLP:conf/www/PezoaRSUV16,DBLP:conf/pods/BourhisRSV17}.

We formalize {\jsonsch} specifications as follows.
First of all, 
we add a parameter $i\in\Nat$ to the semantic function $\semcai{S}{\E}$, 
and we give a definition of $\semcai{\RRef{x}}{\E}$ that is inductive on $i$:
$$\begin{array}{llllllll}
 \semcapar{\RRef{x}}{\E}{0}  &= & \emptyset  
 \qquad\qquad\qquad
 \semcapar{\RRef{x}}{\E}{i+1}  &= & \semcai{\E(x)}{\E}\\[\NL]
\end{array}
$$

This provides a definition of $\semcai{S}{\E}$ that is inductive on the lexicographic pair $(i,|S|)$.
Because of negation, this sequence of interpretations is 
not necessarily monotonic in $i$, but we can still extract an exists-forall limit from it, by stipulating that an instance 
$J$ belongs to the limit $\semca{S}{\E}$ if an $i$ exists such that $J$ belongs to every interpretation 
that comes after $i$:
$$
\semca{S}{\E} = \bigcup_{i\in \mathbb{N}}\bigcap_{j \geq i} \semcapar{S}{\E}{j}
$$

Now, it is easy to prove that this interpretation satisfies {\jsonsch} specifications, since, for guarded schemas,
it enjoys the properties expressed in Theorem~\ref{theo:bigone}, stated below.

\begin{definition}\label{def:closing}
An environment $\E = \Def{x_1}{S_1} , \ldots, \Def{x_{n}}{S_{n}}$ is \emph{closing} for $S$ if all variables used
in the bodies $S_1,\ldots,S_n$ and in $S$ are included in $x_1,\ldots,x_n$. 
\end{definition}

\begin{lemmaxrep}
[Stability]\label{lem:stability}
For every $(S,\E)$ where $\E$ is guarded and closing for $S$, for every $J$: 
$$\begin{array}{llll}
\exists i.\ 
(\forall j \geq i.\ \J \in\semcapar{S}{\E}{j})
\Or
(\forall j \geq i.\ \J \not\in\semcapar{S}{\E}{j})
\end{array}$$
\end{lemmaxrep}

\begin{proof}
We define the degree $d(S)$ of a schema $S$ in $\E$ as follows.
If $S$ is a variable $x$, then $d(x) = d(E(x))+1$.
If $S$ is not a variable, then $d(x)$ is the maximum degree of all unguarded variables in $\E(x)$ and,
if it contains no unguarded variable, then $d(S)=0$.
This definition is well-founded thanks to the guardedness condition.
We prove the lemma by induction on $(J,d(S),S)$, in this order of significance.
We want to prove that, for every triple $(\J,S,\E)$ (but we will ignore $\E$ for simplicity), 
exists a ``fixing index'', that is an $i$ such that,
for any $ j \geq i$, the question whether $\J$ belongs to $\semcapar{S}{\E}{j}$ always yields the same answer.\\[\NL]

Let $S=x$. We want to prove that, for any $\J$:
\\[\NL]$\exists i.\ 
(\forall j \geq i.\ \J \in\semcapar{x}{\E}{j})
\Or
(\forall j \geq i.\ \J \not\in\semcapar{x}{\E}{j})$
\\[\NL]
This is equivalent to the following statement:
\\[\NL]$\exists i.\ 
(\forall j \geq i.\ \J \in\semcapar{\E(x)}{\E}{j-1})
\Or
(\forall j \geq i.\ \J \not\in\semcapar{\E(x)}{\E}{j-1})$
\\[\NL]i.e.\ $\exists i.\ 
(\forall j \geq i.\ \J \in\semcapar{\E(x)}{\E}{j})
\Or
(\forall j \geq i.\ \J \not\in\semcapar{\E(x)}{\E}{j})$
\\[\NL]
 This last statement holds by induction, since $d(x)=d(\E(x))+1$,
hence the term $\J$ is the same but the degree of $e(x)$ is strictly smaller than that of $S=x$.\\[\NL]

Let $S=S'\And S''$. We want to prove that, for any $\J$:
\\[\NL]$\exists i.\ 
(\forall j \geq i.\ \J \in\semcapar{S'\And S''}{\E}{j})
\Or
(\forall j \geq i.\ \J \not\in\semcapar{S'\And S''}{\E}{j})$
\\[\NL]
By induction, for the same $J$ the following statements hold:
\\[\NL]$\exists i.\ 
(\forall j \geq i.\ \J \in\semcapar{S'}{\E}{j})
\Or
(\forall j \geq i.\ \J \not\in\semcapar{S'}{\E}{j})$
\\[\NL]$\exists i.\ 
(\forall j \geq i.\ \J \in\semcapar{S''}{\E}{j})
\Or
(\forall j \geq i.\ \J \not\in\semcapar{S''}{\E}{j})$
\\[\NL]
If we take a witness $I'$ for $i$ in the first property and a witness $I''$ for $i$ in the second property, we have that 
$\max(I',I'')$ is a fixing index for $\J$ and $S$. We reason in the same way for the other boolean operators.\\[\NL]

\hide{
Let $S=\Not S'$. We want to prove that
\\[\NL]$\exists i.\ 
(\forall j \geq i.\ \J \in\semcapar{\Not S'}{\E}{j})
\Or
(\forall j \geq i.\ \J \not\in\semcapar{\Not S'}{\E}{j})$
\\[\NL]
This is equivalent to the following statement, which holds by induction on $S$.
\\[\NL]$\exists i.\ 
(\forall j \geq i.\ \J \not\in\semcapar{S'}{\E}{j})
\Or
(\forall j \geq i.\ \J \in\semcapar{S'}{\E}{j})$
\\[\NL]}

Let $S =\Ite{S_1 \itdots  S_n}{S_{n+1}}$. We want to prove that
\\[\NL]$\exists i.\ 
(\forall j \geq i.\ \J \in\semcapar{\Ite{S_1 \itdots  S_n}{S_{n+1}}}{\E}{j})
\Or
(\forall j \geq i.\ \J \not\in\semcapar{\Ite{S_1 \itdots  S_n}{S_{n+1}}}{\E}{j})$
\\[\NL]
Consider the semantics of $\Ite{S_1 \itdots  S_n}{S_{n+1}}$:
\\[\NL]$\{ \J \M  \J =    [\J_1, \ldots, \J_m], l \in \SetTo{m}    \Rightarrow\ \\
\mbox{\ }\qquad\qquad( ( \forall k \in \SetTo{n}.\ l=k \Rightarrow J_l\in \semcapar{S_k}{\E}{j} )  
\And (l > n \Rightarrow J_l\in \semcapar{S_{n+1}}{\E}{j})\ ) \}$\\[\NL]
The problem $\J\in\semcapar{\Ite{S_1 \itdots  S_n}{S_{n+1}}}{\E}{j}$
gets fixed once we fix this problem for all pairs $(\J_l,S_k)$ and $(J_l,S_{n+1})$ that appear in the definition.
Since each $\J_l$ is a strict subterm of $\J$, each pair has a fixing index by induction, hence the maximum among these
indexes fixes all pairs, hence it fixes the entire question over $(\J,S)$. 
Observe that the fact that each $\J_l$ is \emph{strictly} smaller than $\J$
is essential since,
in general, the degree of each $S_k$ may be bigger than the degree of $S$, since they are all in guarded
position.\\[\NL]

All other guarding operators can be treated in the same way.
\end{proof}

\begin{theoxrep}\label{theo:bigone}
For any $\E$ guarded, the following equality holds:
$$ \semca{\E(x)}{\E} = \semca{x}{\E}
$$
Moreover, 
for each equivalence in Figure \ref{fig:fullsem-ws}, 
the equivalence still
holds if we substitute every occurrence of $\semcai{S}{\E}$ with $\semca{S}{\E}$, obtaining for example:\\[\NL]
$\begin{array}{lllll}
  \semca{\Nam(S)}{\E} 
      &\ =\ \setst{\ J}{  & J= \{ k_1 : J_1,\ldots, k_m : J_m\},  l \in \SetTo{m} \Rightarrow k_{l} \in \semca{S}{\E}\ \ \ } \\
\end{array}$ \\[\NL]
from\\[\NL]
$\begin{array}{lllll}
  \semcai{\Nam(S)}{\E} 
      &\ =\ \setst{\ J}{  & J= \{ k_1 : J_1,\ldots, k_m : J_m\},  l \in \SetTo{m} \Rightarrow k_{l} \in \semcai{S}{\E}\ \ \ }. 
\end{array}$
\end{theoxrep}

\begin{proof}
We want to prove that 

$$
\bigcup_{i\in \mathbb{N}}\bigcap_{j \geq i} \semcapar{x}{\E}{j}=
\bigcup_{i\in \mathbb{N}}\bigcap_{j \geq i} \semcapar{\E(x)}{\E}{j}
$$
Assume that $J\in\bigcup_{i\in \mathbb{N}}\bigcap_{j \geq i} \semcapar{x}{\E}{j}$.
Then, \\$\exists i. \forall j \geq i. J\in  \semcapar{x}{\E}{j}$.
Let $I$ be one $i$ with that property. We have that 
\\$\forall j \geq I. J\in  \semcapar{x}{\E}{j}$, i.e.,
\\$\forall j \geq I. J\in  \semcapar{\E(x)}{\E}{j-1}$, which implies that
\\$\forall j \geq I. J\in  \semcapar{\E(x)}{\E}{j}$, hence
\\$\exists i. \forall j \geq i. J\in  \semcapar{\E(x)}{\E}{j}$.
\\In the other direction, assume 
$J\in\bigcup_{i\in \mathbb{N}}\bigcap_{j \geq i} \semcapar{\E(x)}{\E}{j}$.
Hence, 
\\$\exists i. \forall j \geq i. J\in  \semcapar{\E(x)}{\E}{j}$.
Let $I$ be one $i$ with that property. We have that 
\\$\forall j \geq I. J\in  \semcapar{\E(x)}{\E}{j}$, i.e.,
\\$\forall j \geq I. J\in  \semcapar{x}{\E}{j+1}$, i.e.,
\\$\forall j \geq (I+1). J\in  \semcapar{x}{\E}{j}$, i.e.,
\\$\exists i. \forall j \geq i. J\in  \semcapar{x}{\E}{j}$.

For the second property, assume $J\in \semca{\Nam(S)}{\E}$. We have:
\\$J\in \semca{\Nam(S)}{\E}\ \Iff$
\\$\exists i. \forall j \geq i. J\in \semcapar{\Nam(S)}{\E}{j}\ \Iff$
\\$\exists i. \forall j \geq i. J\in \setst{ J}{  J= \{ k_1 : J_1,\ldots, k_m : J_m\}, l \in \SetTo{m} \Rightarrow k_{l} \in \semcapar{S}{\E}{j}}\ \Iff$
\\$J\in \setst{ J}{  J= \{ k_1 : J_1,\ldots, k_m : J_m\}, l \in \SetTo{m} \Rightarrow 
\exists i. \forall j \geq i. k_{l} \in \semcapar{S}{\E}{j}}\ \Iff$
\\$J\in \setst{ J}{  J= \{ k_1 : J_1,\ldots, k_m : J_m\}, l \in \SetTo{m} \Rightarrow 
k_{l} \in \semca{S}{\E}}$

Assume $J\in \semca{\Ite{S_1 \itdots  S_n}{S_{n+1}}}{\E}$. We have:
\\$J\in\semca{\Ite{S_1 \itdots  S_n}{S_{n+1}}}{\E}\ \Iff$
\\[\NL]$\exists i. \forall j \geq i. J\in \semcapar{\Ite{S_1 \itdots  S_n}{S_{n+1}}}{\E}{j}\ \Iff$
\\[\NL]$\exists i. \forall j \geq i.  J =    [\J_1, \ldots, \J_m], l \in \SetTo{m}    \Rightarrow\ \\
\mbox{\ }\qquad\qquad( \forall k \in \SetTo{n}.\ l=k \Rightarrow J_l\in \semcapar{S_k}{\E}{j} )  
\And (l > n \Rightarrow J_l\in \semcapar{S_{n+1}}{\E}{j})\ )\Iff (*) $
\\[\NL]$(J =    [\J_1, \ldots, \J_m], l \in \SetTo{m}    \Rightarrow\ \\
\mbox{\ }\quad( \forall k \in \SetTo{n}.\ l=k \Rightarrow \exists i. \forall j \geq i.  J_l\in \semcapar{S_k}{\E}{j} ) \\
\mbox{\ }\qquad\And (l > n \Rightarrow \exists i. \forall j \geq i.  J_l\in  \semcapar{S_{n+1}}{\E}{j})\ )\Iff$
\\[\NL]$(J =    [\J_1, \ldots, \J_m], l \in \SetTo{m}    \Rightarrow\ \\
\mbox{\ }\quad( \forall k \in \SetTo{n}.\ l=k \Rightarrow  J_l\in \semca{S_k}{\E} )
\And (l > n \Rightarrow  J_l\in  \semca{S_{n+1}}{\E})\ )$
\\[\NL]The step (*) is not obvious. Here, we have a double implication with the following structure:
\\$\exists i. \forall j \geq i.\ (Q_1 \And \ldots \And Q_n) 
 \Iff (*) (\exists i. \forall j \geq i. Q_1) \And \ldots \And (\exists i. \forall j \geq i. Q_n )$
\\In the $\Implies$ direction, the implication is immediate. In the $\RevImplies$ direction,
for each $\exists i. \forall j \geq i. Q_n$ we may have a different witness $I_1,\ldots,I_n$ for
each existential quantification, but we can choose the maximum $I_M$, since, if $j \geq I_M$,
then $j \geq I_i$ for every $i\in\SetTo{n}$.

The cases for $\Props$ and $\And$ are analogous. In the cases for $\Cont{S}$ and $\Or$, instead, we
have to commute $\exists i. \forall j \geq i.$ with  a disjunction. For example,
assume $J\in \semca{\Cont{S}}{\E}$. We have:
\\[\NL]$J\in\semca{\Cont{S}}{\E}\ \Iff$
\\[\NL]$\exists i. \forall j \geq i.\ J\in \semcapar{\Cont{S}}{\E}{j}\ \Iff $
\\[\NL]$
\exists i. \forall j \geq i.\ J =    [\J_1, \ldots, \J_m] \Implies \exists l \in \SetTo{m}.
                 J_l\in \semcapar{S}{\E}{j}   \Iff (*) $
\\[\NL]$
J =    [\J_1, \ldots, \J_m] \Implies \exists l \in \SetTo{m}.\exists i. \forall j \geq i.\ J_l\in \semcapar{S}{\E}{j}   \Iff $
\\[\NL]$
J =    [\J_1, \ldots, \J_m] \Implies \exists l \in \SetTo{m}. J_l\in \semca{S}{\E} $
\\[\NL]
Here, the $\Iff (*)$ equivalence has this structure:
\\$\exists i. \forall j \geq i.\ (Q_1 \Or \ldots \Or Q_n) 
 \Iff (*) (\exists i. \forall j \geq i. Q_1) \Or \ldots \Or (\exists i. \forall j \geq i. Q_n )$
\\In this case, both directions are immediate.

Finally, assume $J\in \semca{\Not S}{\E}$
\\$J\in\semca{\Not S}{\E}\ \Iff$
\\[\NL]$\exists i. \forall j \geq i.\ J\in \semcapar{\Not S}{\E}{j}\ \Iff$
\\[\NL]$\exists i. \forall j \geq i.\ J\not \in \semcapar{S}{\E}{j}\ \Iff (*)$
\\[\NL]$\forall i. \exists j \geq i.\ J\not\in\semcapar{S}{\E}{j}\ \Iff$
\\[\NL]$\Not(\exists i. \forall j \geq i.\ \J\in\semcapar{S}{\E}{j})\ \Iff$
$J\not\in\semca{S}{\E}\ \Iff$
\\[\NL] For the crucial $\Iff (*)$ step, 
the direction $\Implies$ is immediate. For the direction $\RevImplies$ we use the stability Lemma \ref{lem:stability}.
The stability lemma specifies that the problem $\J\not\in\semcapar{S}{\E}{j}$ becomes definitively true or false after 
a certain index.
If we assume that $\forall i. \exists j \geq i.\ J\not\in\semcapar{S}{\E}{j}$, then
$J\not\in\semcapar{S}{\E}{j}$ holds for infinitely many values of $j$, hence, by the stability Lemma  \ref{lem:stability}, it has a fixing index.
\end{proof}

\hide{
\begin{remark}
When recursion is unguarded, this semantics has the problem that, for a variable such as
$\Def{x}{x}$, its interpretation depends on the arbitrary choice of 
$\semcapar{\RRef{x}}{\E}{0}=\emptyset$. 
Since we restrict our attention to guarded recursion, the problem disappears: indeed,  it would be easy to prove
that, when recursion is guarded, the value of $\semca{\RRef{x}}{\E}$ does not change if we set 
$\semcapar{\RRef{x}}{\E}{0}= K$ for any arbitrary set $K$
\end{remark}
}

From now on, we will assume that,
for any schema document $S\ \Defs(\E)$, recursion in $\E$ is guarded, and that $\E$ is closing for $S$, and we will make the same assumption when discussing the semantics of a schema $S$ with respect to an environment~$\E$. 

The full formal definition of the semantics is in Figure \ref{fig:fullsem-ws}.
$\rlan{r}$ are the strings matched by $r$,
$\semt(\Obj)$ are the {\json} terms whose type is \qobject, 
and similarly for the other types.

\begin{figure*}[!htb]
\small
\setlength{\NL}{0.3ex}
$$
\begin{array}{lllll}

 \semcai{\Type(\Null/\Bool/\Str)}{\E}
   & = & \semt(\Null)/\semt(\Bool)/\semt(\Str)\\[\NL]
 \semcai{\Type(\Arr/\Obj/\Num)}{\E}
   & = & \semt(\Arr)/\semt(\Obj)/\semt(\Num)\\[\NL]
\semcai{\Type(T_1,\ldots,T_{n+1})}{\E} & = &\semcai{\Type(T_1)}{\E} \cup\ldots\cup \semcai{\Type(T_{n+1})}{\E}\\[\NL]
  \semcai{\Mof(q)}{\E} & = & \setst{ J}{ J \in \semt(\Num) \Rightarrow  \exists k \textit{ integer  with } J = k*q} \\[\NL]
  \semcai{\Con(J)}{\E} & = &  \Set{J}  \\[\NL]
  \semcai{\Enu(J_1,\ldots,J_{n+1})}{\E} &=& \semcai{\Con(J_1)}{\E} \cup\ldots\cup \semcai{\Con(J_{n+1})}{\E} \\[\NL]

\semcai{\Len_l^{j} }{\E} &=& \setst{J}{J \in \semt(\Str) \Rightarrow l \leq  \textit{length}(J) \leq j } \\
   \semcai{\Pat(r)}{\E} & = & \setst{J}{J \in \semt(\Str) \Rightarrow J \in  \rlan{r}} \\
   \semcai{\Bet_{m}^{M}}{\E} & = & \setst{J}{ J \in \semt(\Num) \Rightarrow  m \leq J \leq M } \\
   \semcai{\XBet_{m}^{M}}{\E} & = & \setst{J}{ J \in \semt(\Num) \Rightarrow  m < J < M } \\
  \semcai{\Pro_l^j}{\E} &=&\setst{J}{J\in \semt(\Obj) \Rightarrow l \leq \noa{J} \leq j}
 \\[\NL]
 \semcai{\Req(k_1,\ldots,k_n)}{\E} &=&\setst{J}{J\in \semt(\Obj) \Rightarrow \forall k\in \{k_1,\ldots,k_n\}. \exists J'.
     (k:J') \in J }
 \\[\NL]



\semcai{\CIte_{l}^j }{\E} &=&        \{ J \ |  \ J\in \semt(\Arr) \Rightarrow l \leq |J|\leq j  \} \\[\NL]
\semcai{ \Uni }{\E} & = & \setst{J}{ J =  [\J_1, \ldots, \J_n ] \Rightarrow  \forall l,j \in \SetTo{n}. \ l\neq j \ \Implies \ J_l\ \neq J_j}\\[\NL]
  \semcai{\True}{\E} &=&  \setst{J}{J \textit{ is a {\json} value}} \\[\NL]
  \semcai{ \False }{\E}&=& \emptyset \\[\NL]
%
%
%



  \semcai{\Nam(S)}{\E} 
       &=&\setst{ J}{  J= \{ k_1 : J_1,\ldots, k_m : J_m\}, l \in \SetTo{m} \Rightarrow k_l \in \semcai{S}{\E}} \\[\NL] 

%
                                                  
  \semcai{\Props(r_1:S_1,\ldots, r_n:S_n;S) }{\E}&=&
  \{ J \ |  \  J= \{ k_1 : J_1,\ldots, k_m : J_m\}, l \in \SetTo{m} \Rightarrow \   \\[\NL]
               
   &&\quad\quad\quad  
                        ( \forall j \in \SetTo{n}.\ k_l\in  \rlan{r_j}  \Rightarrow J_l\in  \semcai{S_j}{\E})  \wedge  \\
                        & & \quad\quad\quad (k_l\not\in  \rlan{(\key{r_1}|\ldots|\key{r_n})}  \Rightarrow J_l\in  \semcai{S}{\E})\ \}\\[\NL]
                                                  
%
 \semcai{\Ite{S_1 \itdots  S_n}{S_{n+1}}}{\E}
         &=&   \{ J \ |    J=      [\J_1, \ldots, \J_m], l \in \SetTo{m}    \Rightarrow   \\
         &&\quad\quad\quad ( \forall j \in \SetTo{n}.\ l=j \Rightarrow J_l\in \semcai{S_j}{\E}      )   \wedge  \\
          &&\quad\quad\quad   (l > n \Rightarrow J_l\in \semcai{S_{n+1}}{\E})\}  \\[\NL]

  \semcai{\Cont{S}}{\E} &=& \setst{J}{  J=      [\J_1, \ldots, \J_m] \Rightarrow \exists l \in \SetTo{m}.\ J_l\in \semcai{S}{\E} }\\[\NL]
 \semcai{S_1 \And S_2}{\E} & = &  \semcai{S_1 }{\E}  \cap   \semcai{ S_2}{\E} \\
\semcai{\Not S}{\E}  & = & \setst{J}{ J \textit{ is a {\json} value}, J \not\in \semcai{S}{\E}  } \\
  \semcai{ S_1 \vee S_2 }{\E} & = &  \semcai{S_1 }{\E}  \cup  \semcai{ S_2}{\E}\\[\NL]
  \semcai{ S_1 \Implies S_2 }{\E} & = & \semcai{\Not S_1 \Or S_2 }{\E} \\[\NL]
  \semcai{( S_1 \Implies S_2 \ | \ S_3 )}{\E} & = &
  \semcai{(S_1 \And S_2) \Or ((\Not S_1) \And S_3)}{\E} \\[\NL]
  \semcai{\XOr(S_1,\ldots,S_n) }{\E}& = &
  \semcai{\BigOr_{1 \leq l \leq n}({\Not S_1}  \And\ldots \And {\Not S_{l-1}} \And{S_l} \And {\Not S_{l+1}} \And\ldots \And\Not {S_n})}{\E}\\[\NL]
\semcai{\{S_1,\ldots,S_n\} }{\E}&=& \semcai{S_1 \And \ldots \And S_n}{\E}\\[\NL]
 \semcapar{\RRef{x}}{\E}{0}  &= & \emptyset\ \ \ \ \ \ (\text{any arbitrary set of {\json} values could be used}) \\[\NL]
 \semcapar{\RRef{x}}{\E}{i+1}  &= & \semcai{\E(x)}{\E}\\[\NL]
 \semca{S}{\E} &=& \bigcup_{i\in N}\bigcap_{j \geq i}\semcapar{S}{\E}{j} \\[\NL]
\semca{S\ \Defs(\Def{x_1}{S_1} , \ldots, \Def{x_{n}}{S_{n}})}{} &=&
 \semca{S}{\Def{x_1}{S_1} , \ldots, \Def{x_{n}}{S_{n}}}  
\end{array}
$$
\caption{Semantics of the algebra.} 
\label{fig:fullsem-ws}
\end{figure*}




\gcomment{\begin{remark}

To understand the definition of $\PReq$ one may consider the following equivalence, which is
equivalent to the definition, and means that, in order to violate the schema
$\PReq(r_1,\ldots,r_n)$, an instance must be an object and must satisfy 
$ r_i : \False$ for some $i$. An object satisfies $ r_i : \False$ only if no name matches $r_i$, 
hence $( r_1 : \False ) \Or \ldots \Or ( r_n : \False )$ is true for an object iff exists $i$ such that no name of the object matches $r_i$.\footnote{We may, alternatively, define 
$\PReq(r) = \Obj \Implies \Not \Nam(\Not \Pat(r))$.
}

$
\begin{array}{lllll}
\Not  \PReq(r_1,\ldots,r_n) &= \Not (\Obj \Implies
( \Not ( r_1 : \False ) \And \ldots \And \Not ( r_n : \False ) )) \\
&= 
\Obj \And ( ( r_1 : \False ) \Or \ldots \Or ( r_n : \False ) )
\end{array}
$

The definition of $\APReq(r_1,\ldots,r_n)$ is just a reduction to $\PReq(\NotPPPP(r_1|\ldots|r_n))$.
\end{remark}
}

\begin{toappendix} 
\subsection{$\Nam(S)$  encoded through $\POfS(S,\E)$} \label{sec:names}

The assertion $\Nam(S)$ requires that, if the instance is an object, every member name satisfies $S$, which is equivalent
to saying that no member name exists that violates $S$.
Hence, if we translate $S$ into a pattern $r=\POfS(S)$ that exactly describes the strings that satisfy~$S$,
we can translate $\Nam(S)$ into $\CProp{\POfS(\Not S)}{\False}$, which
means:  if the instance is an object, it cannot contain any member whose name matches the complement of $\POfS(S)$.

\newcommand{\EE}{\E}

In order to translate a schema $S$ into a pattern we actually need an environment $\EE$ to associate a definition to 
any variable that appears in $S$; hence, rather than a function $\POfS(S)$, we will define a function $\POfS(S,\EE)$.

We now show how to transform every schema $S$ into a pattern $\POfS(S,\EE)$ such that the following equivalences hold,
where $S \Iff_\EE S'$ means that 
$\semca{S}{\EE}=\semca{S'}{\EE}$.
$$
\begin{array}{llll}
\Type(\Str) \And S & \Iff_\EE & \Type(\Str) \And \Pat(\POfS(S,\EE)) \\
\Nam(S) & \Iff_\EE & \CProp{\POfS(\Not S,\EE)}{\False}
\end{array}
$$

We have already introduced the notations $\NotP(r)$ for the complement of $r$.
We also
define the following abbreviations, where $\TrueP$ matches any string, $\FalseP$ matches no string, and
$r \AndP r'$ matches $\rlan{r}\cap \rlan{r'}$.
$$
\begin{array}{llllllllllllll}
\TrueP & = & .* & \qquad \qquad & \FalseP & = &  \NotP(\TrueP)\\
r \AndP r' & = & \NotP(\NotP(r) \ |\ \NotP(r')) 
\end{array}
$$

Once we have these abbreviations, we can proceed as follows.

For all the ITAs $S$ whose type is not $\Str$, such as $\Mof(q)$, we define $\POfS(S,\EE) = \TrueP$, since they are satisfied by
any string.


For the other operators, $\POfS(S,\EE)$ is defined as follows.
Observe that, while $\POfS(\Mof(q),\EE) = \TrueP$ since $\Mof(q)$ is an Implicative Typed Operator,
$\POfS(\Type(\Num),\EE) = \FalseP$, since $\Type(\Num)$ is not conditional, and is not satisfied by any
string.
Since $\POfS(S,\EE)$ does not analyze the schemas that are nested inside typed operators, 
the definition below is well-founded in presence of guarded recursion: after we have
expanded a variable $x$ once, in the result of any further expansion $x$ will always be guarded, hence we 
will not need to expand it again.
For the operators not cited,
such as $\Len$ and $\Enu$, and the derived boolean operators, we first translate them
into the core algebra, and then we apply the rules below.
$$
\begin{array}{llll}
\POfS(\Type(T),\EE) &=&  \FalseP  & \text{if\ } T\neq \Str\\[\NL]
\POfS(\Type(Str),\EE) &=&  \TrueP  \\[\NL]
\POfS(\Con(J),\EE)  &=& \FalseP & \text{if the type of } J\text{ is not }\Str \\[\NL]
\POfS(\Con(J),\EE)  &=&  \keykey{J} & \text{if the type of } J\text{ is }\Str \\[\NL]
\POfS(S_1 \And S_2,\EE) &=&  \multicolumn{2}{l}{\POfS(S_1,\EE) \AndP \POfS(S_2,\EE)}\\[\NL]
\POfS(\Not S,\EE)  &=&  \multicolumn{2}{l}{\NotP(\POfS(S,\EE))} \\[\NL]
\POfS(\Pat(r),\EE) &=& r \\[\NL]
\POfS(\RRef{x},\EE) &=&  \multicolumn{2}{l}{\POfS(\EE(x),\EE)}  \\[\NL]
\end{array}
$$

It is easy to prove that we have the following equivalences, which allow us to translate
 $\Nam$  into the core algebra.

\begin{property}
For any assertion $S$ and for any environment $\EE$ which maps variables into assertions in a way that 
respects the guarded recursion constraint, the following equivalences hold.
$$
\begin{array}{llll}
\Type(\Str) \And S & \Iff_{\EE} & \Type(\Str) \And \Pat(\POfS(S,\EE)) \\[\NL]
\Nam(S) & \Iff_{\EE} & \CProp{\POfS(\Not S,\EE)}{\False}
\end{array}
$$
\end{property}

\end{toappendix}




\hide{

\section{From {\jsonsch} to the algebraic form}

For the readers who do not know {\json}  Schema, the first important thing to know is that, semantically, {\jsonsch} is reflected
by the algebra: it is a logical formalism characterized by the presence of a rich set of operators --- conjunction, disjunction, negation, 
implication and others --- by the presence of recursive definitions, and by a rich set of assertions on the base types, all having 
the conditional semantics that we described.
As for the syntax, {\jsonsch} assertions are represented as {\json}  objects where, typically, the member name is the
logical operator name and the member value is the parameter. All the different members in an object are connected by conjunction,
so that the algebraic expression:
$$
\Len_3^{\Inf} \And \Pat(\NN a)
$$
is written in {\jsonsch} syntax as:
\begin{quote}
\{ \QQ\xminL\QQ\ : 3, \QQ\xpatt\QQ\ : \QQ$\NN a$\QQ \}
\end{quote}
When the argument of an operator is a number or a string, it is represented as in the previous example. When it
is a set of key-value pairs, as in $\Props(r_1 : S_1,\ldots,r_n : S_n; )$, it is represented
as a {\json}  object, and, when it is a list, it is represented as a {\json}  array.
The {\jsonsch} mechanisms to define recursive variables have been discussed in Section \ref{sec:defsandrefs}.}

\hide{
Our algebra is essentially an algebraic version of {\jsonsch}, where we merge in single operators all the non-algebraic families of operators,
that is, those families of operators whose semantics 
depends on the occurrence of other operators in the same schema object.

The non-algebraic {\jsonsch} families are:
\begin{enumerate}
\item {\xaddProps} acquires its semantics by interacting with the co-occurring assertions \xprops\ and \xpattProps, since it specifies an assertion associated to all members whose keys do not match those matched by the
  two related operators;
\item {\xaddIts} interacts with co-occurring \xit; \xit\ may accept a schema $S$ or an array $S_1,\ldots,S_n$ 
   as argument; in the first case, it is equivalent to $\Ite{}{S}$, and a co-occurring \xaddIts\ is ignored, while in the second
   case it is equivalent to $\Ite{S_1 \itdots  S_n}{\True}$ unless an {\xaddIts} assertion is present, then 
   the assertion of \xaddIts\  substitutes the $\True$ in the tail;
\item the operators {\xif}, {\xthen}, and {\xelse}, when found in the same object, interact in order to define an
  if-then-else operator 
\hide{ \item in {\VerNine}, {\xminC} and {\xmaxC} interact with the co-occurring {\xcont} assertion:
  they give a minimal and a maximal upper bound to the number of elements of the instance array which satisfy the argument
  of {\xcont}.}
\end{enumerate}}

\hide{
The translation from {\jsonsch} to the algebraic form is straightforward and is defined by the following steps.

\smallskip
\noindent
\textbf{Definition normalization.}
For every subschema $S$ that is referred by a 
\QQ\xdref\QQ : \QQ\emph{path}\QQ\ 
operator, we copy $S$ in the \xdefs\ section of the schema, under a name
$f({\mathit{path}})$, where $f$ transforms the path into a flat string, 
and we substitute all references \QQ\xdref\QQ : \QQ\emph{path}\QQ\ with a normalized
reference
\QQ\xdref\QQ : \QQ\#/\xdefs/$f({\mathit{path}})$\QQ, with the only exception of the root path $\QQ \# \QQ$
that is not affected.
At this point, the resulting document
is translated as follows, where $\Tr{S}$ is the translation of $S$,
\key{xroot} is a fresh variable, and where each occurrence of 
``\QQ\xdref\QQ : \QQ\#/\xdefs/$x$\QQ'' is translated as ``$\RRef{x}$''
and each occurrence of
``\QQ\xdref\QQ : \QQ\#\QQ'' is translated as ``$\RRef{\key{xroot}}$'':
$$\begin{array}{llll}
\ensuremath{\langle}\{ &\QQ a_1\QQ : S_1, \ldots, \QQ a_n\QQ : S_n, \QQ\xdefs\QQ : \{ \QQ x_1\QQ : S'_1, \ldots, \QQ x_m\QQ : S'_m \} \;
\} \ensuremath{\rangle}\ =\\[\NL]
& \key{xroot} \ \ \Defs(\Def{xroot} {\Tr{  \{ \QQ a_1\QQ : S_1, \ldots, \QQ a_n\QQ : S_n \}}} ,
\Def{x_1}{\Tr{S'_1}} ,  \ldots, \Def{x_n}{ \Tr{S'_n}})
\end{array}$$
}

\hide{ 
\begin{example}
We consider the following {\JS} document

\begin{Verbatim}[fontsize=\small,xleftmargin=5mm]
{ "properties": {
    "Country": { "type": "string" },
    "City":    { "$ref": "#/properties/Country" } }
}
\end{Verbatim}
Definition normalization produces the following, equivalent schema:
\begin{Verbatim}[fontsize=\small,xleftmargin=1mm]
{"properties": {
   "Country": {"type": "string" },
   "City":    {"$ref": "#/definitions/properties_Country"}},
 "definitions": {"properties_Country": {"type": "string" }}
}
\end{Verbatim}
Which is translated as:
$$\begin{array}{llll}
\RRef{xroot}\ \Defs (&\RRef{xroot} : \Props(\!\!\!&\kw{Country} : \Type(\Str),  \kw{City} : \RRef{properties\_Country};\ \ {\True}),\qquad\qquad \\
 & \multicolumn{2}{l}{
 \RRef{properties\_Country} : \Type(\Str)
           \ \  )
}
\end{array}$$
\end{example}
}


\hide{
\begin{remark}
One may wonder why \xdepS\ and \xdepR\ are translated differently.
The schema
\xdepS : \{ \QQ a\QQ : $S$ \} 		
is translated as  \mbox{$ (\TObj \And \Req(a)) \Implies \Tr{S}$}, since \xdepS\ is an
CTA, hence it is satisfied by any instance that is not an object, which is expressed by the
$\TObj$ hypothesis.
On the other side, when we translate \xdepR, we do not need to add an explicit 
$\TObj$ hypothesis, since the thesis $\Req(r_1,\ldots,r_n)$ is in any case satisfied by any 
instance that is not an object. Of course, adding $\TObj$ to the hypothesis would not 
change the result, but just be redundant.
\end{remark}
}

\hide{
All the other constructs of {\jsonsch}
have an immediately corresponding operator in the algebra, as reported in Table \ref{fig:reftrans}.
The inverse translation from the algebra to {\jsonsch} is immediate.
}

\begin{toappendix}

\section{Translating from {\jsonsch} to the algebraic form}\label{sec:translation}

To translate from {\jsonsch} to the algebra, we first normalize \qdref\ references, as follows.      
For every subschema $S$ that is referred by a 
\QQ\xdref\QQ : \QQ\emph{path}\QQ\ 
operator, we copy $S$ in the \xdefs\ section of the schema, under a name
$f({\mathit{path}})$, where $f$ transforms the path into a flat string, 
and we substitute all references \QQ\xdref\QQ : \QQ\emph{path}\QQ\ with a normalized
reference
\QQ\xdref\QQ : \QQ\#/\xdefs/$f({\mathit{path}})$\QQ, with the only exception of the root path $\QQ \# \QQ$
that is not affected.
At this point, the resulting document
is translated as follows, where $\Tr{S}$ is the translation of $S$,
\key{xroot} is a fresh variable, and where each occurrence of 
``\QQ\xdref\QQ : \QQ\#/\xdefs/$x$\QQ'' is translated as ``$\RRef{x}$''
and each occurrence of
``\QQ\xdref\QQ : \QQ\#\QQ'' is translated as ``$\RRef{\key{xroot}}$'':
$$\begin{array}{llll}
\ensuremath{\langle}\{ &\QQ a_1\QQ : S_1, \ldots, \QQ a_n\QQ : S_n, \QQ\xdefs\QQ : \{ \QQ x_1\QQ : S'_1, \ldots, \QQ x_m\QQ : S'_m \} \;
\} \ensuremath{\rangle}\ =\\[\NL]
& \key{xroot} \ \ \Defs(\Def{xroot} {\Tr{  \{ \QQ a_1\QQ : S_1, \ldots, \QQ a_n\QQ : S_n \}}} ,
\Def{x_1}{\Tr{S'_1}} ,  \ldots, \Def{x_n}{ \Tr{S'_n}})
\end{array}$$

After definition normalization, we translate any assertion $\QQ a\QQ : S$ into the corresponding algebraic operator,
as reported in Table \ref{fig:reftrans}.

\begin{figure*}[!tbp]
\centering
\small
\begin{tabular}{ll}
\toprule
\{ $G_1$, \ldots, $G_n$ \} 
& $\{ \Tr{G_1}, \ldots, \Tr{G_n} \}$ \\
\rowcolor{LightCyan}
\QQ\xall\QQ : [ $S_1$, \ldots $S_n$ ] 
& $\And(\Tr{S_1}, \ldots \Tr{S_n})$   \\
\QQ\xany\QQ : [ $S_1$, \ldots $S_n$ ] 
& $\Or(\Tr{S_1}, \ldots \Tr{S_n})$ \\
\rowcolor{LightCyan}
\QQ\xone\QQ : [ $S_1$, \ldots $S_n$ ] 
& $\XOr(\Tr{S_1}, \ldots \Tr{S_n})$ \\
\QQ\xnot\QQ : $S$  
& $\Not \Tr{S}$ \\
\rowcolor{LightCyan}
\QQ\xif\QQ : $S_{1}$, ``\xthen'' : $S_{2}$, ``\xelse'' : $S_{3}$ 
 & $\Tr{S_{1}} \Implies \Tr{S_{2}} \Else \Tr{S_{3}}$ \\
 \QQ\xconst\QQ : $J$
& $\Con(J)$ \\ 
\rowcolor{LightCyan}
  \QQ\xenum\QQ : $[J_1,\ldots,J_n]$
& $\Enu(J_1,\ldots,J_n)$ \\ 
\QQ\xtype\QQ : \QQ boolean\QQ / \QQ null\QQ / \QQ number\QQ / 
& $\Type(\Bool/ \Null/ \Num/ $  \\
\qquad\quad\; \QQ string\QQ / \QQ array\QQ / \QQ object\QQ
& \quad\quad $\; \Str/ \Arr/ \Obj)$ \\
\rowcolor{LightCyan}
\QQ\xtype\QQ : ``integer''  
& $\Type(\Num) \And \Mof(1) $  \\
%
%
\rowcolor{LightCyan}
 \QQ\xmin\QQ : m 
& $\Bet_m^{\Inf} $\\
\QQ\xmax\QQ : M
& $\Bet_{-\Inf}^M $\\
\rowcolor{LightCyan}
\QQ\xexmin\QQ : m 
& $\XBet_m^{\Inf} $\\
\QQ\xexmax\QQ : M
& $\XBet_{-\Inf}^M $\\
\rowcolor{LightCyan}
\QQ\xmof\QQ : q
& $\Mof(q)$\\
\QQ\xminL\QQ : m 
& $\Len_m^{\Inf}$\\
\rowcolor{LightCyan}
\QQ\xmaxL\QQ : M 
& $\Len_{0}^M$\\
\QQ\xpatt\QQ: r 
 &  $\Pat(r)$ \\
 \rowcolor{LightCyan}
 \QQ\xuniqIt\QQ : \QQ\xtrue\QQ
&  $\Uni$\\
\QQ\xuniqIt\QQ : \QQ\xfalse\QQ
&  $\True$\\
\rowcolor{LightCyan}
\QQ\xminIt\QQ : m
&  $\CIte_m^{\Inf}  $\\
\QQ\xmaxIt\QQ : M
&  $\CIte_0^M  $\\
\rowcolor{LightCyan}
\QQ\xcont\QQ : $S$ 
& $\Cont{\Tr{S}}$ \\ 
\QQ\xit\QQ : [$S_1$, \ldots, $S_n$], 
\QQ\xaddIts\QQ : $S'$
&  $\Ite{\Tr{S_1} \itdots  \Tr{S_n}}{ \Tr{S'}}$\\
\rowcolor{LightCyan}
\QQ\xit\QQ : $S$
&  $\Ite{}{\Tr{S}} $\\
\QQ\xminP\QQ : m 
&  $\Pro_m^{\Inf}$\\
\rowcolor{LightCyan}
\QQ\xmaxP\QQ : M
&  $\Pro_0^M$\\
\QQ\xpropN\QQ : $S$
&  $\Nam({\Tr{S}})$\\
\rowcolor{LightCyan}
\QQ\xreq\QQ : [ \key{k_1},\ldots,\key{k_n} ]
&  $\Req( \key{k_1},\ldots,\key{k_n} ) $\\[\NL]
\QQ\xprops\QQ : \{ i=1..n  \key{k_i} : $S_i$ \},
& $ \Props( i=1..n\ \ \keykey{k_i} : \Tr{S_i}$,    \\
\quad\QQ\xpattProps\QQ : \{i=1..m \key{r_i} :\,$PS_i$ \},
&  \ \ \ \qquad $  i=1..m\ \ \key{r_i} : \Tr{PS_i};$  
\\
\quad\QQ\xaddProps\QQ : $S$
& \ \ \ \qquad$\Tr{S})  $   \qquad\qquad \\
%
\rowcolor{LightCyan}
\QQ\xdepS\QQ : \{ \key{k_1} : $S_1$,\ldots,\key{k_n} : $S_n$ \}
&  $  ((\TObj\And\Req(k_1)) \Implies \Tr{S_1})\And$\\
\rowcolor{LightCyan}
&$\ldots\And ( (\TObj\And\Req(k_n)) \Implies \Tr{S_n}  ) $\\[\NL]
\QQ\xdepR\QQ :
&  $  ( \Req(k_1) \Implies \Req( \key{r^1_1}\ldots,\key{r^1_{m_1}}) )\And$\\
 \qquad   \{  \key{k_1} : [\key{r^1_1}\ldots,\key{r^1_{m_1}}],\ldots,  \key{k_n} : [\key{r^n_1}\ldots,\key{r^n_{m_n}}] \}
&$\ldots\And (\Req(k_n) \Implies \Req(\key{r^n_1}\ldots,\key{r^n_{m_n}}))$\\[\NL]
\rowcolor{LightCyan}
\QQ\xdeps\QQ : \key{obj}
&see two previous cases\\[\NL]
\key{k_1} : $S_1$,\ldots,\key{k_m} : $S_m$, 
&  
$ \key{xroot}\ \Defs(\Def{xroot} {\Tr{  \{ \key{k_1} : S_1,\ldots,\key{k_m} : S_m \}}} ,$\\
\qquad \QQ\xdefs\QQ : \{ \key{x_1} : $S'_1$, \ldots, \key{x_n} : $S'_n$\}
& $ \qquad\qquad \Def{x_1}{\Tr{S'_1}} ,  \ldots, \Def{x_n}{ \Tr{S'_n}})$\\
\rowcolor{LightCyan}
\QQ\xdref\QQ : \QQ\#/\xdefs/$x$\QQ &   $\RRef{x}$ \\
\QQ\xdref\QQ : \QQ\#\QQ &   $\RRef{\key{xroot}}$ \\
\bottomrule
\end{tabular}
\caption{Translation from {\jsonsch} to the algebra.} 
\label{fig:reftrans}
\end{figure*}

\end{toappendix}


%
\hide{Most {\jsonsch} documents only use references to refer either to the entire document of to a schema collected in the
local \xdefs\ section.\GG{We should cite here our database?}
 In order to translate documents that refer to other subschemas of to remote documents,
we first copy the referred subschema in the local \xdefs\ section, and we then apply the rule above.
This mechanism entails some code duplication, but, according to our experiments, this is not a practical problem.
\noindent
For a more detailed analysis of the reference mechanism, see Section \ref{sec:defsandrefs}.
}




%% file: negationclosure.tex

\newcommand{\PIf}{\key{If0}}
\newcommand{\bm}{\key{bm}}
\newcommand{\BM}{\mathcal{B}^{n}}

\subsection{{\jsonsch} is almost negation-closed, but not exactly}

We say that a logic is negation-closed if, for every formula, there exists an equivalent one
where no negation operator appears.
In our algebra, negation operators include $\Implies$, $( S_1 \Implies S_2 \ | \ S_3 )$ and $\XOr$,
since $\Not S$ can be also expressed as $S \Implies \False$ or as $\XOr(S,\True)$.
Negation-closure is usually obtained by coupling each algebraic operator with a dual operator
that is used to push negation inside the first one.
We are going to prove here that negation-closure is ``almost'' true for {\jsonsch} but not completely,
and we are going to exactly describe the situations where negation cannot be pushed through 
{\jsonsch} operators.


According to our collection of GitHub JSON Schema documents, the most common usage 
patterns for $\Props(r_1 : S_1, \ldots, r_n : S_n ; S_a)$ are those where each 
$r_i$ is the pattern $\keykey{k_i}$ that only matches the string $k_i$,\footnote{
we use $\keykey{k}$ to denote the pattern $\NN k' \$$, where $k'$ is obtained from $k$ by escaping all special characters.
}
generated by the use of the {\jsonsch} operator \qprops, and where $S_a$ is either 
$\True$ or $\False$. In these two cases, negation can be pushed through $\Props$ as described
by  Property~\ref{pro:notprop}.


\newcommand{\LN}{\ceil{log_2(n)}}
\newcommand{\Ee}{\mathcal{E}}


We first show how to express $\Not S$
in the negation free algebra when $S=\Props(\keykey{k_1} : \True, \ldots, \keykey{k_n} : \True ; \False)$.
The idea is to define a list of assertions $U_u$, for $u\in\SetTo{n}$, such that  
$\{k'_1:\J'_1,\ldots,k'_m:\J'_m\}\in\semca{U_{u}}{\E}$ iff
$|K' \cap K|\leq u$,
where $K'=\Set{k'_1,\ldots,k'_m}$ and $K=\Set{k_1,\ldots,k_n}$:
$U_u$ puts a bound $u$ on the number of names of $J'$ that are also in $K$ (i.e., in $S$).
At this point, $J'$ violates $S$ iff it has a name that is not in $K$, that is,
iff there exists $i$ such that $J'\in\semca{U_i \And \Pro_{i+1}^{\Inf}}{\E}$: $\J'$ has 
at most $i$ names that are in $K$, but it has more than $i$ fields.

In order to express $U_u$ in a succinct way, we recursively split $[1,\ldots,n]$ into halves
(we actually split $[1,\ldots,2^{\LN}]$)
and, for each \emph{halving interval} $I=[i+1,\ldots,i+(2^l)]$ that we get, we define a variable $U_{I,u}$ that specifies
that $|K' \cap \{ k_{i+1},\dots,k_{i+(2^l)} \}| \leq u$,
and we express $U_{(I_1\cup I_2),u}$ by a combination of $U_{I_1,i}$ and $U_{I_2,j}$.
Formally, let $I(l,p)$ denote the interval of integers
$[(p-1)*(2^l)+1,p*(2^l)]$, that is, the $p$-th interval of length $2^l$, counting from
1, e.g., $I(3,1)=[1,\ldots,8]$, $I(3,2)=[9,\ldots,16]$,\ldots. 
We define now an environment $\Ee(\Set{k_1,\ldots,k_n})$ that contains
a set of variables $U_{l,p,u}$
 such that
 $\{k'_1:\J'_1,\ldots,k'_m:\J'_m\}\in\semca{U_{l,p,u}}{\E}$ 
iff $| K' \cap \setst{k_i}{i\in I(l,p)}| \leq u$.

One variable $U_{(l,p,u)}$ is defined for each {halving interval} $I(l,p)$ that
is included in $[1,2^{\LN}]$, and for each $u\leq 2^l-1$, since $I(l,p)$ contains
$2^l$ elements, hence any constraint with $u \geq 2^l$ would be trivially true.
In the first two lines, we deal with intervals of length $2^0=1$.
The third line exploits the equality 
$I(l,p) = I(l-1,2p-1) \cup I(l-1,2p)$, and it says that 
$| K' \cap K(l,p)| \leq u$
holds if there exists an $i$ such that 
$| K' \cap K(l-1,2p-1)| \leq i$
and
$| K' \cap K(l-1,2p)| \leq (u-i)$.
$$
\begin{array}{lllllll}
\multicolumn{3}{l}{
\Ee(\Set{k_1,\ldots,k_n}) =
} \\
( U_{0,p,0} &:& \Props(k_p : \False; \True)  &  
   1 \leq p \leq n    \\
\ U_{0,p,0} &:& \True  &  
   n+1 \leq p \leq 2^{\LN}    \\
\ U_{l,p,u} 
    &:&\BigOr_{0\leq i \leq u} ( U_{l-1,2p-1,i} \And  U_{l-1,2p,u-i} ) &
    1\leq l \leq \LN,\  \\ 
    & & & 1 \leq p \leq 2^{\LN-l}, \ \\
    &&&   0\leq u \leq 2^{l}-1 \\
\ U_{l,p,u\leq 2^{l-1}} 
    &:&\BigOr_{0 \leq i \leq u } ( U_{l-1,2p-1,i} \And  U_{l-1,2p,u-i} ) &
    1\leq l \leq \LN,\ \\
    & & & 1 \leq p \leq 2^{\LN-l}, \ \\
    &&&   0\leq u \leq 2^{l}-1 \\
\ U_{l,p,u > 2^{l-1}} 
    &:&\BigOr_{u-2^{l-1} \leq i \leq 2^{l-1}} ( U_{l-1,2p-1,i} \And  U_{l-1,2p,u-i} ) &
    1\leq l \leq \LN,\ \\
    & & &  1 \leq p \leq 2^{\LN-l}, \ \\
    &&&   0\leq u \leq 2^{l}-1 \\
    )
\end{array}
$$
The instances of the third line need a number of symbols (computed in Appendix) that grows like $O(n^2)$:
$\Sigma_{l\in\SetTo{\LN}}\Sigma_{p\in\SetTo{2^{\LN-l}}}
   \Sigma_{u\in\SetFromTo{0}{2^l-1}}(4\times (u+1)+1)$.

\begin{toappendix}
 \subsection{Dimension of $\Ee(\Set{k_1,\ldots,k_n})$}
 Computation of $\Sigma_{l\in\SetTo{\LN}}\Sigma_{p\in\SetTo{2^{\LN-l}}}
   \Sigma_{u\in\SetFromTo{0}{2^l-1}}(4\times (u+1)+1)$:
 $$
 \begin{array}{llllllllll}
\Sigma_{l\in\SetTo{\LN}}\Sigma_{p\in\SetTo{2^{\LN-l}}} \Sigma_{u\in\SetFromTo{0}{2^l-1}}(4\times (u+1)+1)\\
=\Sigma_{l\in\SetTo{\LN}}\Sigma_{p\in\SetTo{2^{\LN-l}}} O(2^{2l})\\
=\Sigma_{l\in\SetTo{\LN}}(2^{\LN-l}) \cdot O(2^{2l})\\
=\Sigma_{l\in\SetTo{\LN}} O(2^{\LN+l})\\
=\Sigma_{l\in\SetTo{\LN}} O(2^{2\cdot\LN}) = O(n^2)\\
\end{array}
$$
\end{toappendix}
   
The interval $I(\LN,1)$ includes $\SetTo{n}$, hence an object $J'$ violates 
$\Props(\keykey{k_1} : \True, \ldots, \keykey{k_n} : \True ; \False)$
iff it satisfies $U_{\LN,1,i}$ for some $i$, hence it contains at most
$i$ of the names in $\Set{k_1,\ldots,k_n}$, and it also satisfies $\Pro_{i+1}^{\Inf}$,
hence it contains some extra names.
This construction, based on the counting operator $\Pro_i^j$, allows us to push negation 
through $\Props$ when $S_a=\False$, as shown in Property \ref{pro:notprop},
cases (2) and (3).

\begin{property}[Negation of common use case for $\Props$]\label{pro:notprop}
$$
\begin{array}{llll}
(1)&(\Not \Props(\keykey{k_1} : S_1, \ldots, \keykey{k_n} : S_n ; \True ),\E) \\ 
& \qquad = (\TObj \And \BigOr_{i \in \SetTo{n}}( \Req(\key{k_i}) \And \Props(\keykey{k_i} : \Not S_i; \True ) )
,\E) \\[\NL]
(2)&(\Not \Props(\keykey{k_1} : \True, \ldots, \keykey{k_n} : \True ; \False ),\E)  \\
& \qquad =  
     (\Type(\Obj) \And ( \BigOr_{0 \leq i \leq n} ( U_{\ceil{\log_{2}n},1,i} \And \Pro_{i+1}^{\Inf}) ) 
     , E \cup \Ee(\Set{k_1,\ldots,k_n}) )\\[\NL]
(3)&(\Not \Props(\keykey{k_1} : S_1, \ldots, \keykey{k_n} : S_n ; \False ),\E) \\
& \qquad =  
(\Not \Props(\keykey{k_1} : S_1, \ldots, \keykey{k_n} : S_n ; \True ) \ \Or 
\Not \Props(\keykey{k_1} : \True, \ldots, \keykey{k_n} : \True ; \False ), \E) 
\end{array}
$$
\end{property}

Case (1) shows that, when each $r_i$ has shape $\keykey{k_i}$ and when $S_a=\True$,
then $\Req$ acts as a negation dual for $\Props$. 
Case (3) shows that also in the second most-common use case negation can be 
pushed through
$\Props$, although at the price of a complex encoding, where $\Pro_i^{\Inf}$ plays a crucial role.

A natural question is which other use cases can be expressed, maybe through more and more complex encodings.
To answer this question, we first introduce a bit of notation.

\newcommand{\Ss}{\mathcal{S}_S}
\newcommand{\SEq}[1]{[#1]_S}

\begin{notation}
Given an assertion $S=\Props(r_1 : S_1, \ldots, r_n : S_n ; S_a)$ and a string $k$,
the functions $\SEq{k}$ and $\Ss(k)$ are defined as follows.
\begin{compactenum}
\item $\SEq{k} = \setst{k'}{\forall i\in \SetTo{n}.\ k\in\rlan{r_i}\Iff k'\in\rlan{r_i}}$:
   the set of strings that match exactly the same patterns as $k$.
\item $\Ss(k)$: let $I = \setst{i}{k\in\rlan{r_i}}$;
     if $I=\emptyset$ then $\Ss(k)=S_a$ else $\Ss(k) = \And_{i\in I} S_i$:
    the conjunction of the schemas that must be satisfied by $J'$ if $k:J'$ is a member of an object
       that satisfies $S$.
\end{compactenum}
\end{notation}

\hideforspace{
The function $\Ss$ is an alternative way of representing $\Props(r_1 : S_1, \ldots, r_n : S_n ; S_a)$,
since it enjoys the following property.

\begin{property}
Given an assertion $S=\Props(r_1 : S_1, \ldots, r_n : S_n ; S_a)$, 
for any object $J$, 
$\J$ belongs to $\semca{S}{\E}$ iff, for every member $k:J'$ of $J$,
we have that $\J'\in\semca{\Ss(k)}{\E}$.
\end{property}
}

We now prove that Property \ref{pro:notprop} exhausts all cases where  
$\Not\Props(r_1 : S_1, \ldots, r_n : S_n ; S_a)$ can be expressed without negation.
When we say $(S,\E)$ can be expressed as $(S',\E')$, this means that
$\semca{S}{\E} = \semca{S'}{\E'}$ -- proof in Appendix.

\begin{theoxrep}\label{the:not}
Given $S=\Props(r_1 : S_1, \ldots, r_n : S_n ; S_a)$,
if exist $k_1, k_2$ such that (1) $\SEq{k_1}$ and $\SEq{k_2}$ are both infinite, and 
(2) exist both $\J^{+}_1$ and $\J^{-}_2$ such that 
$\J^{+}_1\in\semca{\Ss(k_1)}{\E}$ and $\J^{-}_2\not\in\semca{\Ss(k_2)}{\E}$,
then $(\Not S,\E)$ cannot be expressed without negation.
\end{theoxrep}

\begin{appendixproof}
$\Not\Props(r_1 : S_1, \ldots, r_n : S_n ; S_a)$ is satisfied by every instance that is an object 
and such that it contains at least one member $k:J'$ such that 
$\J'\not\in\semca{\Ss(k)}{\E}$.

Assume that $k_1, k_2, \J^{+}_1, \J^{-}_2$ 
exist, and assume that a positive
$D = S_0\ \Defs(\E')$ with $\E'=\Def{x_1}{S'_1},\ldots,\Def{x_n}{S'_n}$ expresses 
$(\Not\Props(r_1 : S_1, \ldots, r_n : S_n ; S_a),\E)$, in order to reach a contradiction.
Consider a name $k\in\SEq{k_2}$ that does not appear in
any $\Req$ operator that is in $D$: since $\SEq{k_2}$ is infinite, such $k$ exists.
Let $mm$ be a number that is bigger than every lower bound $m$ that appears in any 
$\Pro_m^M$ in $D$
and strictly bigger than the number of fields of any object found inside any $\Con$ or $\Enu$ operator in $D$.
Consider a set of $mm$ different names $\Set{k'_1,\ldots,k'_{mm}}$ that belong to $\SEq{k_1}$ --- such a set exists since $\SEq{k_1}$ is infinite. Now consider the following two objects:
$$
\begin{array}{llll}
\{ \QQ k'_1 \QQ : \J^{+}_1, \ldots, \QQ k'_{mm} \QQ : \J^{+}_1, \QQ k \QQ : \J^{-}_2 \} & \qquad\qquad O_1 \\
\{ \QQ k'_1 \QQ : \J^{+}_1, \ldots, \QQ k'_{mm} \QQ : \J^{+}_1 \} & \qquad\qquad O_2 \\
\end{array}$$

A generic $S'$ satisfies \emph{OneImpliesTwo} if
$O_1\in\semcai{S'}{\E'}\Implies O_2\in\semcai{S'}{\E'}$.
We now
prove that every assertion $S'$ inside $D$ satisfies  \emph{OneImpliesTwo},
by induction on the lexicographic pair $(i,|S'|)$, where $|S'|$ is the size of $S'$.
In this way, we prove that $D$ satisfies \emph{OneImpliesTwo}, which is a contradiction 
since $\Not S$ is satisfied by $O_1$, thanks to the
$\QQ k \QQ : \J^{-}_2$ member, while $\Not S$ is not satisfied by $O_2$.

Every assertion that cannot distinguish two objects satisfies \emph{OneImpliesTwo}.
$\Con$ and $\Enu$ assertions in $D$ do not contain $O_1$ or $O_2$ since these are too big, 
by construction.
The $\Props$ and $\Nam$ assertions can only fail because of the presence of a field, never for
its absence, hence they satisfy \emph{OneImpliesTwo}. 
The name $k$ does not appear in any $\Req$ assertion in $D$, hence all $\Req$ assertions
satisfy \emph{OneImpliesTwo}. If  $O_1\in\semcai{\Pro_m^M}{\E'}$, then
$O_2$ satisfies the upper bound since $O_2$ is shorter than $O_1$, and it satisfies the lower bound
since it has $mm$ members, and $mm\geq m$ by construction.
If $O_1\in\semcai{S_1\And  S_2}{\E'}$, then $O_1\in\semcai{S_1}{\E'}$
and $O_1\in\semcai{S_2}{\E'}$, hence the same holds for $O_2$ by induction on the size of
$S$, hence $O_2\in\semcai{S_1\And S_2}{\E'}$.
The same holds if we exchange $\And$ with $\Or$ and \emph{and} with \emph{or}.
For variables, the thesis follows by induction
on $i$, since $\semcapar{x_j}{\E'}{i+1}=\semcai{S_j}{\E'}$ and $S_j$ is a subterm of $D$, 
and \emph{OneImpliesTwo} holds trivially when $i=0$.
Hence $D$ itself enjoys \emph{OneImpliesTwo}, hence $D$ does not express $(\Not S,\E)$.
\end{appendixproof}

\begin{corxrep}\label{pro:nottwo}
Let $S=\Props(r_1 : S_1, \ldots, r_n : S_n ; S_a)$.
The assertion
$(\Not S,\E)$ can be expressed without negation only if $(S,\E)$ can be expressed either as 
$(\Props(\keykey{k_1} : S_1, \ldots, \keykey{k_n} : S_n ; \True ),\E')$ or as 
$(\Props(\keykey{k_1} : S_1, \ldots, \keykey{k_n} : S_n ; \False ),\E')$, for some $\E'$.
\end{corxrep}

\begin{appendixproof}
Assume that $(\Not S,\E)$ can be expressed without negation.
By Theorem \ref{the:not}, it is not the case that exist $k_1$ and $k_2$ 
such that (1) $\SEq{k_1}$ and $\SEq{k_2}$ are both infinite, and 
(2) exist both $\J^{+}_1$ and $\J^{-}_2$ such that 
$\J^{+}_1\in\semca{\Ss(k_1)}{\E}$ and $\J^{-}_2\not\in\semca{\Ss(k_2)}{\E}$.
Hence, either for every $k$ such that $\SEq{k}$ is infinite
there exists no $\J^{+}$ such that $\J^{+}\in\semca{\Ss(k)}{\E}$, hence
$\semca{\Ss(k)}{\E}=\semca{\False}{\E}$, or
for every $k$ such that $\SEq{k}$ is infinite
there exists no $\J^{-}$ such that $\J^{-}\in\semca{\Ss(k)}{\E}$, hence
$\semca{\Ss(k)}{\E}=\semca{\True}{\E}$.
In the first case, $\Props(r_1 : S_1, \ldots, r_n : S_n ; S_a)$
can be expressed as $\Props(\keykey{k_1} : S'_1, \ldots, \keykey{k_m} : S'_m ; \False)$, as follows:
every $k_f$ such that $\SEq{k_f}$ is finite, so that
$\SEq{k_f}=\Set{k'_1,\ldots,k'_l}$, is transformed into a finite set of simple constraints
$\keykey{k'_1} : \Ss(k_f),\ldots,\keykey{k'_l} : \Ss(k_f)$, and the additional constraint $\False$ expresses the fact that
every name $k$ such that $\SEq{k}$ is infinite must satisfy the assertion $\False$.
In the second case, we reason in the same way to prove that the schema can be expressed as
$(\Props(\keykey{k_1} : S'_1, \ldots, \keykey{k_m} : S'_m ; \True),\E)$.
\end{appendixproof}

Theorem \ref{the:not} gives an abstract characterization of
the schemas whose negation cannot be expressed. Observe that $k_1$ and $k_2$ may coincide,
as long as $\Ss(k_1)$ is not trivial, where \emph{$(S,E)$ is trivial} when either $\semca{S}{\E}=\semca{\True}{\E}$, or $\semca{S}{\E}=\semca{\False}{\E}$.
Corollary \ref{pro:nottwo} rephrases the Theorem, hence specifying that Property \ref{pro:notprop} is exhaustive: negation cannot
be pushed through $\Props$ unless the schema is equivalent to one of those presented in 
Property \ref{pro:notprop}. Since these are inexpressibility results, the theorem condition is not syntactic,
but is decidable, and can also be used to derive results at the syntax level, as we are going to show below. 

In terms of the original \xpattProps\ and \xaddProps\ operators, Theorem \ref{the:not} shows that
the negation-free complement of a schema that contains \xpattProps\ at the top level is only expressible when the schema can be rewritten into one where \xpattProps\ is not used.
For a schema $S$ that contains \xaddProps\ at the top level, its complement has a  negation-free expression only if $S$ can be rewritten into one where \xaddProps\ is associated to a trivial schema.

Theorem \ref{the:not} also has  two other interesting corollaries.

\begin{corollary}\label{cor:not}
\begin{compactenum}
\item $(\Not\Nam(S),\E)$ can be expressed without negation if, and only if, either 
$\semca{\{\Type(\Str),S\}}{\E}$ is finite or  $\semca{\{\Type(\Str),\Not S\}}{\E}$
is finite.
\item $(\Not \Props(; S),\E)$  can be expressed without negation if, and only if, $(S,\E)$ is trivial.
\end{compactenum}
\end{corollary}

While $\Nam(S)$ is a universally quantified property ``every name in $\J$ belongs to $S$'',
$\Not\Nam(S)$ specifies that \emph{there exists} a name that satisfies $\Not S$;
Corollary \ref{cor:not}(1) specifies that $\Not\Nam(S)$ has a negation-free expression
only in the finitary cases when either the allowed names, or the forbidden names, form a finite
set, so that $\Nam$ is another operator that does not admit, in general, the negation-free expression
of its negation dual in {\jsonsch}. 

Finally, by definition, $(\Not \Props(; S),\E)$ requires the presence of one field whose value satisfies $S$,
independently of its name; Corollary \ref{cor:not}(2) specifies that this assertion cannot be expressed in
the negation-free fragment of {\jsonsch}, for any non-trivial $(S,\E)$.

This last property indicates a big difference with array operators. Arrays can be described as 
objects where
the field names are integers greater than 1, with the extra constraint that, whenever the field name
$n+1$ is present, with $n\geq 1$, then $n$ must be present as well.
From this viewpoint, Corollary \ref{cor:not}(2) says that, while arrays have a positive operator 
$\Cont{S}$ to require the presence of at least one element that satisfies $S$, objects have no
negation-free way of requiring the presence of such a field.
Despite this crucial difference, the final result is quite similar: for arrays, as happens for objects,
the negation of the fundamental $\IteK$ operator can be expressed ``almost'' always, but with some
precise exceptions. 

We first show how negation can be expressed in the most common cases.

\newcommand{\BACK}{\!\!\!\!\!\!}

\begin{property}[Negation of common use cases for $\IteK$]\label{pro-notitems}
\[
\begin{array}{lllll}
(1) & \Not\Ite{}{S_a} =   \Type(\Arr) \And     \Cont{\Not S_a} & \\[\NL]
(2) &\multicolumn{2}{l}{
\mbox{if for each $i$, $\semca{S_i}{\E}\subseteq\semca{S_a}{\E}$:}} \\
& \Not\Ite{S_1 \itdots S_n}{S_a}   = \\
& \qquad  Type(\Arr)  \And  (\BigOr_{i\in\SetTo{n}}( 
\Ite{\True_1\itdots\True_{i-1},\Not S_i}{S_a} \And  \CIte_i^{\Inf})
\Or \Cont{\Not S_a} ) 
\\[\NL]
(3) &\Not\Ite{S_1 \itdots S_n}{\False} = \\
& \qquad \Type(\Arr) \And   (\BigOr_{i\in\SetTo{n}}( 
     \Ite{\True_1\itdots\True_{i-1},\Not S_i}{\True} \And  \CIte_i^{\Inf}) \Or\  \CIte_{n+1}^{\Inf}) 
 \\[\NL]
\end{array}
\]
\end{property}

Observe that the second case of Property \ref{pro-notitems} includes the standard case when $S_a=\True$, and the third case subsumes the case when any $\semca{S_i}{\E}=\semca{\False}{\E}$
for some $S_i$ since, in that case, $\semca{\Ite{S_1 \itdots S_n}{S_a}}{\E}$ is the same as
$\semca{\Ite{S_1 \itdots S_{i-1}}{\False}}{\E}$.
The three cases above include the quasi-totality of the $\IteK$ assertions that we found in our 
collection. However, they do not include one specific case: the one when $n>0$,
there exists an $i$ where $\semca{S_i}{\E}\not\subseteq\semca{S_a}{\E}$,  
for all $i$ $\semca{S_i}{\E}\neq\semca{\False}{\E}$, 
and $\semca{S_a}{\E}\neq\semca{\False}{\E}$.
In this specific case, negation cannot be expressed.

\begin{theoxrep}\label{the:notitems}
The algebra without negation cannot express 
($\Not \Ite{S_1\itdots S_n}{S},\E$) when $n\neq 0$, 
all schemas $S_1,\ldots,$ $S_n$, $S_a$, are non-empty in $\E$,
and there exists an $i$ in $\SetTo{n}$ and $\J_i^{-}$ 
such that $\J_i^{-}\in\semca{S_i\And\Not S_a}{\E}$.
\end{theoxrep}

\newcommand{\Tn}{\mathit{Tn}}
\newcommand{\Ts}{\mathit{Ts}}
\newcommand{\Ta}{\mathit{Ta}}
\newcommand{\TO}{\mathit{To}}
\newcommand{\Tt}{\mathit{G}}

\begin{appendixproof}
Assume that a positive document 
$D = x_j\ \Defs(\Def{x_1}{S'_1},\ldots,\Def{x_m}{S'_m})$ expresses the assertion
 $S=\Not \Ite{S_1\itdots, S_n}{S_a},\E$,  when $n\neq 0$, 
all schemas $S_1,\ldots,$ $S_n$, $S_a$, are non-empty in $\E$,
and there exists $i$ in $\SetTo{n}$ such that 
$\semca{S_i\And\Not S_a}{\E}\neq\semca{\False}{\E}$.

$\Not \Ite{S_1\itdots S_n}{S_a}$ 
is satisfied by any $\J$ that is an array and  has either an element at
a position $j\leq n$ that satisfies $\Not S_j$, or an element after position $n+1$ (included) that satisfies $\Not S_a$.
Let $nn$ by the maximum among the lengths of the array prefixes and the array constants 
that appear in $S'_j$
and the parameter $n$ of the hypothesis.
Consider the following two arrays, where $\J^{+}\in\semca{S_a}{\E}$ and $\J_i^{-}\in\semca{S_i\And \Not S_a}{\E}$
$$
\begin{array}{lr}
\quad[ J_1,\ldots, J_{i-1}, \J_i^{-}, J_{i+1}, \ldots, J_{nn}, \J^{+}, \J_i^{-} ] \ & \qquad\qquad A_1 \\
\quad[ J_1,\ldots, J_{i-1}, \J_i^{-}, J_{i+1},  \ldots, J_{nn},  \J^{+}, \J^{+} ] \ & \qquad\qquad A_2 \\
\end{array}
$$
In these arrays, all elements $J_1,\ldots,J_{nn}$ are chosen to satisfy the corresponding $S_j$, if their position
$j$ is before $n$, or $S_a$ otherwise, which is possible since all these schemas are not empty. 

A generic $S'$ satisfies \emph{OneImpliesTwo} if
$A_1\in\semcai{S'}{\E'}\Implies A_2\in\semcai{S'}{\E'}$.
We now
prove that every assertion $S'$ inside $D$ satisfies  \emph{OneImpliesTwo},
by induction on the lexicographic pair $(i,|S'|)$, where $|S'|$ is the size of $S'$.
In this way, we prove that $D$ satisfies \emph{OneImpliesTwo}, which is a contradiction 
since $\Not S$ is satisfied by $A_1$, thanks to the last element
$\J_i^{-}$, while $A_2$ does not satisfy $\Not S$.

For variables, boolean expressions, and non-array typed operators we reason as in the proof of
Theorem \ref{the:not}.
$\Con$ and $\Enu$ assertions in $D$ do not contain $A_1$ or $A_2$ since these are too big, 
by construction.
Any $\CIte_m^M$ that is satisfied by the first one holds for the second, since they have the same length.
The same holds for any $\Cont{S_a}$, since the first contains the same elements as the second. 
Both fail $\Uni$, by construction.
Finally, for any $\Ite{S'_1\itdots, S'_m}{S'}$ in $S'$, we know that it accepts $A_1$, hence it accepts
$\J^{+}$ in position $nn+1$, hence it accepts it in position $nn+2$ as well since $nn \geq m$, hence it 
accepts $A_2$. Hence $D$ is not equivalent to $\Not \Ite{S_1\itdots, S_n}{S_a}$.
\end{appendixproof}

Hence, we are again in a situation where negation can be pushed through $\IteK$ in almost all cases
of practical interest, but not always.

Observe that, while $\Req$ and $\ContK$ can express the negation of $\Props$ and $\IteK$ in most cases, but not always,
$\Props$ and $\IteK$ can \emph{always} express the negation of $\Req$ and $\ContK$: 

\begin{property}[Full negation for $\Req$ and $\ContK$]
$$\begin{array}{llll}
\Not\Req(\key{k_1},\ldots,\key{k_n})
&=&
\TObj \And  (\Props(\keykey{k_1} : \False;\True) \Or \ldots \Or\Props(\keykey{k_n} : \False;\True))
\\[\NL]
\Not\Cont{ S}
&=&
\Type(\Arr) \And   \Ite{}{\Not{S}}
 \\[\NL]
\end{array}$$
\end{property}

This quasi-duality can be explained as follows. 
$\Ite{S_1 \itdots S_n}{S}$  is a universal-implicative quantification over the 
elements of an array: for every element, if its position~$i$ is before $n$, then its value satisfies $S_i$; if
it is strictly greater, then its value satisfies $S$.
The assertion $\Cont{S}$ is existential-assertive: there exists one element that satisfies $S$;
the same classification can be used for the pair $\Props$-$\Req$. 
In both cases, the existential-assertive element of the
pair is somehow less expressive than the negation of its universal-implicative companion: $\Req$ lacks the ability
to describe infinite sets of names and the associated schemas, $\ContK$ is not able to distinguish the head and the tail 
of the array.

\hideforspace{
This quasi duality can be described and explained as follows.
Assertion $\Ite{S_1 \itdots S_n}{S}$  is a universal-implicative quantification over the 
elements of an array: for every element, if its position~$i$ is before $n$, then its value satisfies $S_i$; if
it position $i$ is strictly greater than $n$, then its value satisfies $S$.
The assertion $\Cont{S}$ is existential-assertive: there exists one element that satisfies $S$. It is almost the 
existential-assertive dual of $\Ite{S_1 \itdots S_n}{S}$, but is slightly less expressive since it lacks the ability to distinguish positions in the head
from those in the tail (Theorem \ref{the:notitems}).

Similarly, the operator $\Props({r_1} : S_1, \ldots, {r_n} : S_n ; S )$ is a universal-implicative quantification over the 
members of an object: for every member, if the name matches $r_i$, then its value satisfies $S_i$, and if
matches the complement of ${r_1}, \ldots, {r_n}$, then its value satisfies $S$.
The operator $\Req(\key{k_1},\ldots,\key{k_n})$  is a conjunction of assertions that are existential-assertive
over the members: there exists one member that 
matches $\keykey{k_1}$, one that matches $\keykey{k_2}$, and so on. 
It is only a weak dual of universal-implicative  $\Props({r_1} : S_1, \ldots, {r_n} : S_n ; S )$, 
since it lacks the ability to specify a pattern with infinite domain and, in the infinite domain case, it has no way to
relate a name with a specific schema (Theorem \ref{the:not}).
}

We conclude this section with the last two operators whose negation cannot be expressed
in the  language without negation: $\Mof(q)$ and $\Uni$. 
In the next section we will show how negation can be pushed through all the other operators.

\begin{theoxrep}\label{pro:notothers}
The following pairs  cannot be expressed in the algebra without negation:
\begin{compactenum}
\item ($\Not \Mof(q),\E$), for any $q>0$.
\item ($\Not \Uni,\E$).
\end{compactenum}
\end{theoxrep}

\begin{appendixproof}
(1) $\Not \Mof(q)$:  
Assume towards a contradiction that a  positive document 
$D = S_0\ \Defs(\E')$ with $\E'=\Def{x_1}{S_1},\ldots,\Def{x_n}{S_n}$ expresses  $\Not \Mof(q)$. 
Let us choose a number $N$ such that $N>q$ (hence, $N>0$), 
and $N>M$ and $N>m$ for any bound $m$ and $M$, different from $\Inf$, 
that is found in any assertion $\Bet_{m}^M$, $\XBet_{m}^M$ inside $D$.

We say that a generic $S'$ is Full Or Finite (FOF) for $i$ over the closed interval $[N,2N]$, 
if $[N,2N]\cap \semcai{S}{\E'}$ is either equal to $[N,2N]$, 
or is finite.
We prove that any subexpression $S'$ of $D$ is FOF over $[N,2N]$ for any $i$,
by induction on the lexicographic pair $(i,|S'|)$.
For the variables, in the case $i=0$ the empty set is finite, and the inductive step is 
immediate since $\semcapar{x_j}{\E'}{i+1}=\semcai{S_j}{\E'}$ and $S_j$ is a subterm of $D$.
Typed operators whose type is not $\Num$ accept all numbers, hence are Full. An interval operator whose
bounds are both smaller than $N$ has empty intersection with $[N,2N]$, and is Full when $M=\Inf$.
The positive $\Mof(q)$ operator has a finite intersection with every finite interval.
Union and intersection of two subsets of $[N,2N]$ which are
either finite or full is finite or full. Hence, $D$ is FOF over $[N,2N]$ for any $i$.
The limit
$\bigcup_{i\in \Nat}\bigcap_{j \geq i} \semcapar{D}{}{j}$ can be infinite only
if exists $i$ such that $\semcapar{D}{}{j}$ is infinite for $j\geq i$, hence $\semca{D}{}$
is full or finite as well.
However, $\Not \Mof(q)$ is not FOF: it is not full on $[2,2N]$ since the interval contains at least one multiple of $q$,
by $N>q$, and its intersection with $[N,2N]$ is not finite.

(2) Assume that a positive $D$ expresses $\Not \Uni$. Choose an integer $N$ strictly greater than any
$l$ that appears as lower bound in a $\CIte_{l}^{j}$ in $D$ 
and also greater than the length of any array that appears in a $\Con$ or $\Enu$ assertion in $D$.
Define the following two arrays, the first one ending with a repetition of $N$. 
$$\begin{array}{lr}
A_1 = [ 1, 2,\ldots, N-1, N, N] 
\qquad\qquad
A_2 = [ 1, 2,\ldots, N-1, N] 
\end{array}$$
We prove by induction on the lexicographic pair $(i,|S'|)$,
that the semantics $\semcai{S'}{\E}$ of any subexpression $S'$ of $D$ that includes $A_1$ 
includes $A_2$ as well. 
When $S=x$, we prove that by induction on $i$: in the base case, $\semcapar{x}{\E}{i}=\emptyset$
does not contain $A_1$, and when $i=i+1$ hence $\semcapar{x}{\E}{i+1}=\semcai{\E(x)}{\E}$,
we conclude by induction on $i$.
Non-array typed assertion include both $A_1$ and $A_2$.
All $\Con$ and $\Enu$ assertions refuse both, since the arrays that they
enumerate are shorter than $N$.
Since the length $N$ is greater than any lower bound $l$, if $\CIte_{l}^{j}$ accepts the first it also accepts the second.
For any  $\Ite{S_1\itdots S_n}{S}$ assertion, since $A_2$ is an initial subarray of $A_1$, if the assertion is satisfied
by $A_1$, it is also satisfied by $A_2$. 
For $S_1\And S_2$ we conclude by induction on the size, since implication of satisfaction is preserved by $\And$,
and similarly for $\Or$.
Since $A_1$ belongs to $\semca{D}{}$ by assumption, $A_2$ belongs to $\semca{D}{}$,
which contradicts the hypothesis, since $A_2$ satisfies $\Uni$.
\end{appendixproof}

\subsection{Closed algebra}

As we have seen, {\jsonsch} does not enjoy negation-closure, but is endowed with universal-existential
pairs $\Props$/$\Req$ and $\IteK$/$\ContK$ that enjoy an imperfect duality.
We now define a more regular algebra by adding some \emph{negative} operators, 
to obtain a closed algebra
where each operator has a real negation dual, and negation can be fully eliminated.
In our experience, this 
negation-closed algebra is practical both to reason about {\jsonsch} and to implement tools for {\jsonsch} analysis. Indeed, our tool allows the user to transform a schema into its algebraic form and to generate a witness for the input schema \cite{DBLP:conf/edbt/AttoucheBCFGLSS21},
and our witness generation algorithm crucially uses the negation elimination algorithm for the closed algebra,
that we will present now, in order to generate witnesses for schemas that contain negation.

\hide{
The closed algebra complements the algebra with the following seven operators:
$\NotPat$, $\NotMof$,  $\PReq$, $\APReq$, $\ExNam$, $\NotUni$, $\ContAfterK$.
The operators $\APReq$ and $\ExNam$ can be expressed using $\PReq$ with some pattern manipulation,\GG{This is a bit
complicated. We may decide to only add the four that cannot be expressed through pattern manipulation,
plus pattern manipulation.} and
$\NotPat$ can be expressed using $\Pat$ and pattern manipulation, while $\PReq$, $\NotMof$, $\NotUni$
and $\ContAfterK$
cannot be expressed without negation.
In the following grammar, $m$ is any number, while $n$ is an integer with $n\geq 0$.
The notation $\True^1\itdots\True^{n}$ indicates a sequence of $n$ copies of $\True$.
$$
\begin{array}{llll}
\NotPat(r)  & = & \TStr \Implies \Not \Pat(r) \\[\NL]
\NotMof(q)  & = & \TNum \Implies \Not \Mof(q) \\[\NL]
\PReq(r_1:S_1,\ldots,r_n:S_n) & = & \TObj \Implies \BigAnd_{i\in\SetTo{n}} \Not  \Props(r_i:\Not S_i ; \True) \\[\NL]
\APReq((r_1,\ldots,r_n) : S) & = & \TObj \Implies \Not  \Props(r_1:\True,\ldots,r_n:\True ; \Not S) \\[\NL]
\ExNam(S) & = & \TObj \Implies \Not\Nam(\Not S) \\[\NL]
\NotUni & = & \TArr \Implies \Not\Uni \\[\NL]
\ContAfter{n}{S} & = & \TArr \Implies \Not\Ite{\True^1\itdots\True^{n}}{\Not S} \\[\NL]
\end{array}
$$
}

The closed algebra completes the algebra with the following four dual operators:
 $\PReq$,  $\ContAfterK$, $\NotMof$, $\NotUni$, none of which, by the theorems we presented, can be expressed in the algebra without negation.
We also add a pattern complement operator $\NotP(r)$ to the regular expressions language,
that is useful for complexity reasons, since, while regular expressions are closed under complement, the size of the plain regular expression that represents $\NotP(r)$ is, in the worst case,
doubly exponential with respect to the size of $r$.
The semantics of these operators is defined as follows (where the notation $\True^1\itdots\True^{n}$ indicates a sequence of $n$ copies of $\True$).
$$
\begin{array}{llll}
\PReq(r_1:S_1,\ldots,r_n:S_n) & = & \TObj \Implies  \BigAnd_{i\in\SetTo{n}} \Not  \Props(r_i:\Not S_i ; \True) \\[\NL]
\ContAfter{n}{S} & = &  \TArr \Implies \Not\Ite{\True^1\itdots\True^{n}}{\Not S} \\[\NL]
\NotMof(q)  & =\ & \TNum \Implies \Not \Mof(q) \\[\NL]
\NotUni & = & \TArr \Implies \Not\Uni 
\end{array}
$$

The operator $\PReq(r_1:S_1,\ldots,r_n:S_n)$ specifies that, if the instance is an object, then, for
each $i\in \SetTo{n}$, it must possess a member whose name matches $r_i$ and whose value
satisfies $S_i$. It is strictly more expressive than $\Req$, since it allows one to require a name that belongs to an
infinite set $\rlan{r_i}$, and it associates a schema~$S_i$ to each required pattern $r_i$. 
In the closed algebra, we regard $\Req$ as an abbreviated form of $\PReq$ where every pattern
has the shape $\keykey{k}$ and every associated schema is $\True$.

$\ContAfter{n}{S}$ specifies that, if the instance is an array, it must contain at least one element 
that satisfies $S$ in a position that is strictly greater than $n$. This operator has an expressive power that 
is slightly greater than the $\Cont{S}$ operator, since it can distinguish between the head and the tail of the array.
In the closed algebra, we regard $\Cont{S}$ as an abbreviation for $\ContAfter{0}{S}$.

The operators $\NotMof(q)$ and $\NotUni$ are just the duals of $\Mof(q)$ and $\Uni$. 
In the next section we prove that these four operators are all that we need to make {\jsonsch}
negation-closed.

\subsection{Proving negation closure: the not-elimination algorithm}\label{sec:notelim}

\newcommand{\IteAtK}{\akey{itemAt}}
\newcommand{\IteAfterK}{\akey{itemsAfter}}
\newcommand{\IteAt}[2]{\IteAtK({#1}:{#2})}
\newcommand{\IteAfter}[2]{\IteAfterK({#1}:{#2})}

We prove negation closure through the definition of a not-elimination algorithm, which eliminates any instance of 
negation from any expression in the closed algebra. This algorithm starts with a simplification phase,
aimed at reducing the complexity of the following phase.
In this simplification phase we use the following derived operators, similar to those used in JSL for
arrays \cite{DBLP:conf/pods/BourhisRSV17}: 
$$\begin{array}{llll}
\IteAt{i}{S} = \Ite{\True^1\itdots\True^{i-1},S}{\True}
\ \ \ \ \ 
\IteAfter{i}{S} = \Ite{\True^1\itdots\True^{i}}{S}
\end{array}$$ 
These are the simplification steps.
\begin{compactenum}
\item $\IteK$ and $\Props$ simplification: we rewrite each
    $\Ite{S_1,\ldots,S_n}{S_a}$
      as $\IteAt{1}{S_1}\And \ldots\And\IteAt{n}{S_n}\And\IteAfter{n}{S_a}$
      and each
      $\Props(\key{r_1} : S_1,\ldots,  \key{r_n} : S_n; S_a)$
      as $\Props(\key{r_1} : S_1;\True) \And\ldots\And\Props(\key{r_n} : S_n;\True)
             \And \Props(\NotP(\key{r_1}|\ldots|\key{r_n})  : S_a;\True)$.
\item Type simplification: we rewrite each $\Type(T_1,\ldots,T_n)$ as 
    $\Type(T_1)\Or\ldots\Or \Type(T_n)$.
\item Const-elimination: we rewrite every instance of $\Con$ and of $\Enu$,
   with the only notable exception of $\Con(\xtrue)$ and $\Con(\xfalse)$, through the repeated application of the rules shown in Figure~\ref{fig:const}, 
   as also done in~\cite{DBLP:conf/issta/HabibSHP21}.
\item $\Nam$ elimination: we rewrite every instance of $\Nam(S)$ using $\Props( r_S : \False; \True)$,
   as discussed below.
\item Not-explicitation: we rewrite every instance of
$S_1 \Implies S_2$, $( S_1 \Implies S_2 \ | \ S_3 )$ and $\XOr(S_1,\ldots,S_n)$, according
to their definition; the only remaining boolean operators are $\Not$, $\And$, $\Or$,
$\True$,~$\False$. 
\end{compactenum}


\begin{figure*}[t]
\setlength{\NL}{0.2ex}
$$
\begin{array}{lllll}
\Enu(J_1,\ldots,J_n) &=& \Con(J_1) \Or \ldots \Or \Con(J_n) \\[\NL]
\Con(\xnull) & = &  \Type(\Null)   \\[\NL]
\Con(n) & = &  \Type(\Num) \And \Bet_n^n  & n\in \Num 
\\[\NL]
\Con(\key{s)} & = &  \Type(\Str) \And \Pat(\keykey{s})  
\qquad\qquad\qquad\qquad\qquad\qquad\qquad& \key{s}\in \Str 
\\[\NL]
\Con([J_1,\ldots,J_n]) &=&
 \multicolumn{2}{l}{
 \Type(\Arr) \And \CIte_n^n \And \IteAt{1}{\Con(J_1)},\ldots,\IteAt{n}{\Con(J_n)}
 } \\[\NL]
 \multicolumn{4}{l}{
 \Con( \{ \key{k_1} : J_1,\ldots, \key{k_n} : J_n \})
 = \ 
\Type(\Obj) \And \Req(k_1,\ldots,k_n) 
        \And \Pro_0^n }\\[\NL]
&&
\multicolumn{2}{l}{
\qquad\qquad\And\ \Props(\keykey{k_1}:\Con(J_1);\True)
                                         ,\ldots, \Props(\keykey{k_n}:\Con(J_n);\True)}
                                                  \\[\NL]
\end{array}
$$
\caption{Elimination of $\Enu$ and $\Con$.}
\label{fig:const}
\end{figure*}

%
$\Nam(S)$ is eliminated by transforming $S$ into a pattern $r_S$ that matches all and only the strings 
that satisfy $S$, 
and by declaring that only fields whose name matches $r_S$ are allowed, as exemplified here:
$$
\begin{array}{lllll}
\Nam(\Pat(\QQ\NN x\QQ) \Or \Con(\QQ zz\QQ)) = \Props( \QQ\NN x | \NN zz \$\QQ: \True\ ; \False )
\end{array}
$$
This phase is fully specified in the Appendix.

Na\"{\i}f not-explicitation may exponentially increase the size of the input schema,
since the translation of $\XOr(S_1,\ldots,S_n)$ takes $n$ copies of each argument,
and that of $( S_1 \Implies S_2 \ | \ S_3 )$ takes two copies of $S_1$.
This explosion can be easily avoided, by substituting each duplicated argument of these two operators
with a fresh variable,
so that the not-explicitation phase would only multiply the number of occurrences of these variables,
but not the entire subschemas represented.
Moreover, the obvious encoding of $\XOr(x_1,\ldots,x_n)$ produces an expression whose size is in $O(n^2)$,
but there exists an alternative encoding with linear size, that we present in the 
Appendix.
Hence, not-explicitation can be implemented in such a way that its output size is linear in the input size,
and the same holds for the other phases of simplification.\footnote{In our implementation we adopted the basic algorithm, 
having verified that in our schema corpus~\cite{schema_corpus} of more than 80k real-world schemas, $\XOr$ has
on average 2.3 arguments, and most of the time these arguments are extremely
small.} 

\begin{toappendix}
\subsection{Linear encoding of oneOf}  

We describe here a linear-size encoding of $\XOr(S_1,\ldots,S_{2^q})$.
Let $I(l,p)$ denote the interval of integers
$[(p-1)*(2^l)+1,p*(2^l)]$, that is, the $p-th$ interval of length $2^l$, where we count from
1.
For a fixed $q$, $I(l,p)$ is a subinterval of $[1,2^q]$ iff $0\leq l \leq q$ and $1 \leq p \leq 2^{q-l}$,
hence, the total number of subintervals of $[1,2^q]$ that have the shape $I(l,p)$, where $l$ and
$p$ satisfy that condition, is
$
\Sigma_{l \in \SetFromTo{0}{q}}2^{q-l} = 2*(2^q)-1
$, linear in $2^q$.
Given a set of schemas $S_1,\ldots,S_{2^q}$, we can define a set of $2*(2^q)-1$ 
variables $N_{l,p}$, one for each subinterval $I(l,p)$ of $\SetTo{2^q}$, and a set 
of variables $O_{l,p}$ such that:
\begin{enumerate}
\item $N_{l,p}$ is equivalent to the conjunction of $\Not(S_i)$ for all $i\in I(l,p)$,
   hence $N_{l,p}$ is satisfied iff none of these schemas is satisfied;
\item $O_{l,p}$ is satisfied iff one and only one of the schemas indexed by an $i\in I(l,p)$ is satisfied.
\end{enumerate}
These two sets of variables are inductively defined as follows;
observe that an interval $I(l+1,p)$ can be split in two halves as follows:
$I_{l+1,p} = I_{l,2p-1} \cup  I_{l,2p}$.
 $$
\begin{array}{lllllll}
N_{0,p} &=& \Not S_p  &
O_{0,p} &=& S_p  \\[\NL] 
N_{l+1,p} &=& N_{l,2p-1} \And  N_{l,2p}  &
O_{l+1,p} &=& ( O_{l,2p-1} \And  N_{l,2p} ) \Or ( N_{l,2p-1} \And  O_{l,2p} )
\end{array}
$$
 
The size of this environment is linear in $2^q$, and the variable $O_{q,1}$ encodes
$\XOr(x_1,\ldots,x_{2^q})$.

\end{toappendix}

%



On this simplified form, we now apply the two fundamental steps.

\begin{compactenum}
\item Not-completion of variables: this is a key technical step, since not-elimination needs to deal with the
presence of recursive variables. In this step, for every variable $\Def{x_n}{S_n}$ we define a 
complement variable $\Def{not\_x_n}{\Not S_n}$, which will then be used to eliminate negation
applied to $x_n$.
\item Not-pushing: given a not-completed pair $(S,E)$
 we repeatedly push negation inside every $\Not S'$ expression until 
negation reaches the leaves and is removed.
\end{compactenum}

\paragraph*{Not-completion of variables}\label{sec:notcompletion}
Not-completion of variables is a key step that allows us to deal with the combined presence of unrestricted negation
and recursive variables.
In particular, not-completion transforms a set of definitions as follows:
%
$$\begin{array}{lllll}
\text{not-completion}(S \ \Defs(\Def{x_1}{S_1}\ldots, \Def{x_n}{S_n})) = \\[0.8ex]
\ \ S\ 
\Defs(\Def{x_1}{S_1}, \ldots, \Def{x_n}{S_n},
\Def{not\_x_1}{\Not S_1} , \ldots, 
\Def{not\_x_n}{\Not S_n})
\end{array}$$

As a result, every variable $\key{x}$ has a complement variable $\NotVar{x}$ defined in the obvious way:
$\NotVar{x_i}=not\_x_i$ and $\NotVar{not\_ x_i}=x_i$. 
Variable $\NotVar{x}$ will later be used for not-elimination.

\begin{property}
Let 
$(\Def{x_1}{S_1}\ldots, \Def{x_n}{S_n})$
be a closing environment.
Then, for every variable $x_i$ with $i\in\SetTo{n}$, we have:
$$
\begin{array}{llll}
\semca{ \Not x_i\ \Defs(\Def{x_1}{S_1}, \ldots, \Def{x_n}{S_n})}{} \ = \\
\semca{not\_x_i\ \Defs(\Def{x_1}{S_1}, \ldots, \Def{x_n}{S_n},
\Def{not\_x_1}{\Not S_1} , \ldots, 
\Def{not\_x_n}{\Not S_n})}{}
\end{array}
$$
\end{property}

\hide{
\begin{theorem}
Old version: Let us define 
$$\akey{eval}\ S\ \akey{using}\ \ADef{x_1}{S_1},\ldots,\Def{x_n}{S_n}$$
as an abbreviation for 
$$\ADef{\mathit{freshVar}}{S}, \Def{x_1}{S_1},\ldots,\Def{x_n}{S_n}$$
where $\mathit{freshVar}$ is an unused variable name, so that 
$\akey{eval}\ S\ \akey{using}\ S'$ evaluates~$S$ using all variables defined in $S'$,
ignoring the distinction between the root variables and the others.

Let $CS$ be the result of not-completion applied to $S$, where $S$ is $\ADef{x_1}{S_1},\ldots,\Def{x_n}{S_n}$.
Then, for every variable $x_i$ defined in $S$, we have:
$$
\begin{array}{llll}
\semca{\akey{eval}\ {\NotVar{x_i}}\ \akey{using}\ CS}{()} = 
\semca{\akey{eval}\ {\Not \RRef{x_i}}\ \akey{using}\  CS}{()}
\end{array}
$$
\end{theorem}
}

\paragraph*{The not-pushing algorithm}
\gcomment{
Not-elimination uses the six negative operators listed in Section \ref{sec:full}.
The operator $\NotPat(r)$ is actually equivalent to $\Pat(not\ r)$ where $not\ r$ is the pattern that matches everything that is not matched by $r$.
As discussed in section \ref{sec:names} 
applying the same pattern-inversion operation, it is possible to translate both $\Nam(S)$ and $\ExNam(S)$ in $r :: \False$, with a pattern $r$ that depends on $S$.
In the same way, we can translate 
$\APReq((r_1,\ldots,r_n):S) $ into $\PReq(\NotP(r_1|\ldots|r_n):S)$.
Hence, if we consider the possibility of pattern rewriting, the only negative
operators that are really needed for not-elimination are: 
$\NotMof(n)$, $\NotUni$, and $\PReq$. Every other operator can be complemented 
using another positive {\json}  schema operator.

\GG{Suggestion for the implementers: please consider the possibility to 
implement $\Type[T_1,\ldots,T_n]$,
so that the complement of $\Type(T)$ does not really need a disjunction}
}
The not-pushing phase pushes negation down any algebraic expression up to its complete elimination.
Not-pushing is defined by the rules in Figure 
\ref{fig:notelimvar}. 
Observe that the negation of each
conditional operation asserts the corresponding type, while the negation of $\Con$ is actually conditional:
if the value is a boolean, then it is equal to \xfalse/\xtrue.

Not-pushing over $\Len_0^{M}$ or $\Len_i^{\Inf}$  
generates one satisfiable bound and one that is actually illegal 
($\Len_0^{-1}$ or $\Len_{\Inf+1}^{\Inf}$). Rather than splitting the rule in three cases, we just assume that
the illegal bound is eliminated from the resulting disjunction, and that a trivial 
operator $\Len_0^{\Inf}$ is just rewritten as $\True$ before not-pushing.
An analogous assumption is made for
the $\Bet$, $\XBet$, $\Pro$, $\CIte$ operators.

\hide{
$$\begin{array}{llll}
\Len_0^{\Inf} & = & \True \\[\NL]
\Len_n^{m} & = & \TStr \Implies \False & \ \ \ \text{if } n > m \\[\NL]
\Len_{\Inf}^{\Inf} & = & \TStr \Implies \False & \\
\end{array}
$$}

\newcommand{\EQ}{\ =\ }
\renewcommand{\EQ}{\,=\,}
\newcommand{\SEP}{\ ;\ }

\setlength{\SaveNL}{\NL}
\setlength{\NL}{0.6ex}

\begin{figure*}[!tb]
\small
\[
\begin{array}{llll}
\multicolumn{4}{l}{
\Not \True \EQ  \False \SEP
\Not \False \EQ  \True \SEP
\Not (S_1 \And S_2) \EQ  (\Not S_1) \Or (\Not S_2) \SEP
\Not (S_1 \Or S_2) \EQ  (\Not S_1) \And (\Not S_2) \SEP
\Not (\Not S) \EQ  S 
}\\[\NL]
\Not(\Type(T)) & = & \BigOr( \Type(T') \M T' \neq T ) \\[\NL]
\Not(\Con(\xtrue)) &=& \BigOr( \Type(T) \M T \neq \Bool ) \Or \Con(\xfalse) \\[\NL]
\Not(\Con(\xfalse)) &=& \BigOr( \Type(T) \M T \neq \Bool ) \Or \Con(\xtrue) \\[\NL]\Not(\Len_{i}^{j}) & = &
\Type(\Str) \And (\Len_{0}^{i-1} \Or \Len_{j+1}^{\Inf}) \\[\NL]
\Not(\Pat(r)) &=& \Type(\Str) \And \Pat(\NotP(r)) \\[\NL]
\Not(\Bet_{m}^{M}) & = &
\Type(\Num) \And (\XBet_{-\Inf}^{m} \Or \XBet_{M}^{\Inf}) \\[\NL]
\Not(\XBet_{m}^{M}) & = &
\Type(\Num) \And (\Bet_{-\Inf}^{m} \Or \Bet_{M}^{\Inf}) \\[\NL]
\Not(\Mof(q)) & = &
\Type(\Num) \And \NotMof(q)  \\[\NL]
\Not(\NotMof(q))  & = & \TNum \And  \Mof(q) \\[\NL]
\Not(\CIte_i^j) & = &
\Type(\Arr) \And (\CIte_{0}^{i-1} \Or \CIte_{j+1}^{\Inf}) \\[\NL]
\Not(\Uni) & = &
\Type(\Arr) \And \NotUni\\[\NL]
\Not(\NotUni) & = & \TArr \And \Uni \\[\NL]
\Not(\IteAt{i}{S}) &=&
\Type(\Arr) \And \IteAt{i}{\Not S_i} \And \CIte_i^{\Inf}\\[\NL]
\Not(\IteAfter{n}{S}) &=&
\Type(\Arr) \And \ContAfter{n}{\Not S}\\[\NL]
\Not(\ContAfter{n}{S}) & = & \TArr \And \IteAfter{n}{\Not S} \\[\NL]
\Not(\Pro_{i}^{j}) & = & 
\Type(\Obj) \And (\Pro_{0}^{i-1} \Or \Pro_{j+1}^{\Inf}) \\[\NL]
\Not(\Props(\key{r} : S);\True)
 & = &
\Type(\Obj)  \And  \PReq(\key{r} : \Not S)
\\[\NL]
\Not(\PReq(r:S)) & = & \TObj \And \Props(r:\Not S;\True) \\[\NL]

\Not(\RRef{x}) &=& \NotVar{x}\\[\NL]
\end{array}
\]
\caption{Not-pushing rules --- standard rules are collected in the first line.} 
\label{fig:notelimvar}
\end{figure*}

\setlength{\NL}{\SaveNL}



\hide{
Negation of $\Props$ requires the use of negative operators when we have 
non-trivial patterns or a schema for the additional properties that is not $\True$.
The vast majority of objects schema do not make use of $\xpattProps$ and impose
no constraint for the additional properties, and in this case we can have full
not-elimination, with no need of the special negative assertions, as follows. 
$$\begin{array}{llll}
\Not( \Props(\keykey{k_1} : S_1,\ldots, \keykey{k_n} : S_n; \True) ) \\[\NL]
\qquad\qquad= \Type(\Obj) \ \And 
(\BigOr_{i\in 1..n} (\Req(\key{k_i})\And \Props(\keykey{k_i} : \Not S_i;\True) ) )\\[\NL]
\end{array}$$
}

The following property is not difficult to prove.

\hideforspace{
\begin{property}
The not-elimination procedure eliminates all instances of negation.
\end{property}
}
%
%
%

\begin{property}
The not-elimination procedure preserves the semantics of the schema.
\end{property}

If we define the size of $\IteAt{i}{S}$ in the natural way as $1+\log(i)+|S|$, rather than considering the
length of its definition, and do the same for $\IteAfterK$,
then it is easy to see that the output size of not-elimination is linear with respect to the input size.

%% file: examples.tex

Consider the following {\jsonsch} document.

\begin{Verbatim}[fontsize=\small,xleftmargin=5mm]
{ "properties": {"a": {"not": {"$ref": "#"}}} }
\end{Verbatim}

It is quite obscure, and seems to suggest an infinite alternation of \QQ a\QQ\ and its negation.
Yet, not elimination makes its semantics more clear.
We write it in our algebra as follows. 
$$
\begin{array}{lllll}
x\ \Defs(\Def{x}{\Props( \keykey{a} : \Not{x};\True)})
\end{array}
$$

By applying not-completion, we get the following definition 
(for readability, we omit the trivial ``;\True'' at the end
of the $\Props$ operator).
$$
\begin{array}{lllll}
x\ \Defs(\Def{x}{\Props( \keykey{a} : \Not{x})},
\Def{not\_x}{\Not \Props( \keykey{a} : \Not{x}) } )
\end{array}
$$

This is how not-elimination would now proceed within our algebra (we push 
$\Not$ through $\Props$ using Property \ref{pro:notprop} and use $\{\ ,\ \}$ for conjunction):
$$
\begin{array}{lllll}
\Defs(& \ \Def{x}{\Props( \keykey{a} : \Not{x})}, \Def{not\_x}{\Not \Props(  \keykey{a} : \Not{x})}  \ )\ \To \\[0.8ex]
\Defs(& \ \Def{x}{{\Props( \keykey{a} :\NR{x})}}, \Def{not\_x}{ \{\Type(\Obj), \Req(a),  \Props(\keykey{a} :  \Not \Not \RRef{x}) \} } \ )\ \To \\[0.8ex]
\Defs(& \ \Def{x}{{\Props( \keykey{a} :\RRef{not\_x})}}, \Def{not\_x}{ \{\Type(\Obj),  \Req(a),  \Props(\keykey{a} :  \RRef{x})\}}\ ) \\[0.8ex]
\end{array}
$$

We now substitute \key{not\_x} with its definition, and obtain
a much clearer schema: \emph{if} the instance is an object with an \emph{a}
member, then the value of that member \emph{must} be an object with an \emph{a} member,
whose value satisfies the same specification:
$$
\begin{array}{lllll}
x\ \Defs(\ \ \Def{x}{{\Props( \keykey{a} : \{ \Type(\Obj), \Req(a), \Props(\keykey{a} :  \RRef{x})\} )}}\ \ )
\end{array}
$$

These are some examples of values that match that schema:
$$
1, \{\QQ b\QQ : 2 \}, \{\QQ a\QQ : \{ \QQ a\QQ : \QQ foo\QQ \} \} , \{ \QQ a\QQ : \{ \QQ a\QQ : \{ \QQ a\QQ : \{ \QQ a\QQ : \xnull \} \}  \} \} 
$$\DC{NICE, should we say in the conclusion that it is worth exploring not-elimination adoption to improve schema readability/clarity ? }

%% file: core.tex

Many of the operators of our algebra are obviously redundant, and it is natural to look for minimal subsets where the
rest of the algebra can be mapped.

Such a \emph{core algebra} has a theoretical, descriptive, and practical interest. 
It reduces the effort necessary to define the
semantics of the algebra and to carry on any proof that is exhaustive on the syntax. This has been proven true
time and again for the core of relational algebra~\cite{AHV}. Descriptively, our core algebra shows the 
expressive power of {\jsonsch} in a more concise way.
Practically, we are implementing a tool to check satisfiability and inclusion for {\jsonsch}, and we made it simpler by mapping everything to the core.

\smallskip
\noindent
\textbf{Core algebra.}
One possible choice is reported in the next grammar, 
 where all symbols $q$, $m$, $M$, $n$, $i$, $j$ and $r$ have the
same range as in the full grammar.

$$
\begin{array}{llll}
B & ::= &  \xtrue \M \xfalse \\[\NL]
S & ::= &\ 
\Con(B)   \M  \Pat(r) \M \Bet_{m}^{M}  \M \Mof(q) \\[\NL]
&& \M   \Props(r : S; \True)   \M \Pro_{i}^{j} 
   \\[\NL]
&&  \M \Ite{S_1 \itdots  S_n}{S_{n+1}} 
     \M \CIte_i^{\Inf}
     \M \Uni  
     \\[\NL]
&& 
\M  \RRef{x}  \M  S_1 \And S_2 \M \Not S \\[\NL] 
D & ::= & S\ \Defs(\Def{x_1}{S_1} , \ldots, \Def{x_n}{S_n}) \\[\NL]
\end{array}
$$

Let us consider the operators restricted or deleted.
We have seen already how $\Enu$ and $\Con$ can be mapped to boolean constants. $\Len_i^j$ can be expressed
as a pattern, while $\XBet$ can be expressed by $\Bet$ and negation. The operator $\Props$ can be restricted to the unary
form $\Props(r : S; \True)$ thanks to conjunction, and the additional $S$ case in $\Props(r_1 : S_1,\ldots,r_n : S_n; S)$
can be translated as $\Props((\NotP ( r_1|\ldots|r_n )): S; \True)$.
$\Req$ can be expressed by negation of $\Props$.
The translation of $\Nam(S)$ is a bit more complicated, and requires both $\Props$ and some pattern manipulation
(Appendix \ref{sec:names}).
$\Cont{S}$ can be expressed by negation of $\Ite{}{S}$.
The upper bound of $\CIte_i^j$ can be expressed by putting $\False$ at the position $j+1$ of the 
$\Ite{}{}$ operator, and hence $\CIte_i^{\Inf}$ is sufficient. 
Finally, the fact that all boolean operators can be expressed by $\And$ and $\Not$ is not surprising.
As discussed below, the encoding of $\Type(T)$ is not obvious, and relies on $\Con(B)$.

The core algebra enjoys the properties of closure and minimality. 

\begin{property}[closure]
Every expression in the full algebra can be translated into an equivalent expression in the core algebra.
\end{property}

\begin{proofsketch}
We give a translation for each operator of the full algebra. We start with $\Type(T)$.
For each type, as shown in Figure~\ref{fig:translation_core}, we provide the conjunction of two 
conditional typed assertions that
are not compatible, hence are only satisfied by instances of any other type. Hence, their negation
is only satisfied by instances of the corresponding type.
The only exception are the $\Bool$ type, where we use $\Con(B)$ which is the only typed operator that is
not conditional, since it is only satisfied by the specified value, and the $\Null$ type: 
since we have no typed operators for $\Null$,  we take the 
complement of the other five types. Pattern ``$\hat{}\ \$$'' matches the empty string, and
the pattern ``.'' matches every non-empty string.

\begin{figure}[ht]
$$\begin{array}{lll}
\Type(\Str) &=& \Not ( \Pat(\ \hat{}\ \$) \And (\Pat(.) )) \\[\NL]
\Type(\Num) &=& \Not ( \Bet_0^0 \And \Bet_1^{\Inf}  ) \\[\NL]
\Type(\Bool) &=&\Con(\xfalse) \Or \Con(\xtrue) \\[\NL]
\Type(\Obj) &=& \Not ( \Pro_{0}^{0} \And\Pro_{1}^{\Inf} ) \\[\NL]
\Type(\Arr) &=& \Not ( \Ite{\False}{\False} \And \CIte_1^{\Inf}  ) \\[\NL]
\Type(\Null) &=& \Not \Type(\Str) \And \Not \Type(\Num) \And \Not \Type(\Bool)  \And \\[\NL] 
                         & & \Not \Type(\Obj) \And \Not \Type(\Arr)  \\[\NL]
\end{array}
$$

\caption{Translation of types to the core algebra.}\label{fig:translation_core}
\end{figure}

The translation of the other operators has been shortly described in the text. 
%
%
%

\end{proofsketch}

Translation from the full algebra to the core may incur size explosion because of the translation of
$( S_1 \Implies S_2 \ | \ S_3 )$ and $\XOr(S_1,\ldots,S_n)$ (see Section~\ref{sec:notelim}).
If we ignore this issue, translation is linear, with only one exception, as detailed below.

\begin{property}[Linearity]
If we translate an expression in the full algebra that does not contain a $( S_1 \Implies S_2 \ | \ S_3 )$ or 
an $\XOr(S_1,\ldots,S_n)$ subexpression into an equivalent expression in the core algebra,
the result has linear size with respect to the input, with the only exception of the operator
$\CIte_i^j$, whose translation has length $O(j)$, which is exponentially bigger than the representation of $j$.
\end{property}


\begin{property}[Minimality]
The core algebra is minimal, in the sense that no operator in the core algebra can be expressed using the others. 
\end{property}

\smallskip
\noindent
\textbf{Negation-closed core.}
Similarly to the negation-closed algebra, we can define a negation-closed core algebra.
In this algebra, we substitute negation with the seven dual operators listed in the grammar below.
The first three are present in {\jsonsch} ($ \Type(T)$, $S_1 \Or S_2$, and $\XBet_{m}^{M}$),
and the other four ($\PReq(r:S)$, $\ContAfter{n}{S}$, $\NotMof(q)$, and $\NotUni$)
are the unary versions of those added to the {\jsonsch} algebra to make it closed. 

$$
\begin{array}{llll}
B & ::= &  \xtrue \M \xfalse \\[\NL]
S & ::= &\ 
\Con(B)   \M  \Pat(r) \M \Bet_{m}^{M}  \M \Mof(q) \\[\NL]
& &  \M 
     \Props(r : S; \True)   \M \Pro_{i}^{j} 
   \\[\NL]
&&  \M \Ite{S_1 \itdots  S_n}{S_{n+1}} 
     \M \CIte_i^{\Inf}
     \M \Uni   \\[\NL]  
& &  \M \RRef{x}  \M  S_1 \And S_2 \\[\NL] 
    &&  \M \Type(T) \M S_1 \Or S_2  \M \XBet_{m}^{M} \\[\NL]
    && \M \NotMof(q) \M \NotUni \M \PReq(r:S) \\[\NL]
    & &    \M \ContAfter{n}{S} \\[\NL]
D & ::= &  S\ \Defs(\Def{x_1}{S_1} , \ldots, \Def{x_n}{S_n})
\end{array}
$$

%% file: towardsnewver.tex

\subsection{The $\Ex_i^j S$ operator}\label{sec:contains}

{\VerNine} introduced the new operators \xminC: $n$ and \mbox{\xmaxC: $n$}, where $n$ is a natural number,
with the following semantics: consider a schema that contains, at the top level, the three assertions
``\xcont: S,\ \ \xminC: m,\ \ \xmaxC: M''.
An instance $J$ satisfies this combination iff: if it is an array, then it contains at least
$m$ and at most $M$ elements that satisfy $S$. When \xminC\ is missing, its value defaults to 1,
while a missing \xmaxC\ means no upper limit.
We model this by adding an operator $\Ex_l^j S$ to the algebra, with $l\in\Nat$ and
$j\in\Nat^{\Inf}$. The semantics of $\Ex_l^j S$ is defined as follows.
$$\begin{array}{lllll}
\semcai{\Ex_{l}^{j} S}{\E} \ = \ 
\{\, J  \ \ |& \!\! J =  [\J_1, \ldots, \J_n ] \Rightarrow l\ \leq \ |\,\setst{o}{ o\in\SetTo{n} \And 
 \J_o \in \semcai{S}{\E}}\,|\ \leq\ j  \ \}
\end{array}$$

The operator $\Ex_l^j S$  cannot be expressed in the algebra (proof in the Appendix).

\begin{theoxrep}\label{sec:exists}
The pair $(\Ex_l^j S,\E)$  cannot be expressed in the algebra if $(S,\E)$ is not trivial and
either $l\geq 2$ or $j\neq\Inf$.
\end{theoxrep}

\begin{appendixproof}
Assume that $(S,\E)$ is not trivial and that $D = S_0\ \Defs{(\E')}$ expresses $(\Ex_2^{\Inf} S,\E)$.
Consider $J_1\in\semca{S}{\E}$ and $J_0\not\in\semca{S}{\E}$.
If we say that $n$ is the head-length of an operator $\Ite{S_1\itdots S_n}{S'}$,
let $N$ be the maximum among the head-lengths of all instances of this operator inside $D$ 
and the lengths of all arrays that appear inside the arguments of $\Enu$ and $\Con$ in $D$.
Consider two arrays $A_1$ and $A_2$ of length $N+3$, starting with $N+1$ copies of $J_0$:
$$A_1 = [J_0,\ldots,J_0,J_1,J_1], \qquad\qquad A_2 = [J_0,\ldots,J_0,J_1,J_0]$$
%
%
Only $A_1$ should belong to $\semca{D}{}$, but we can prove
by induction on $i$, and on the size of $S$ when $i$ is equal, that, for every subterm $S'$ of $D$,
$A_1\in\semcai{S'}{\E'}$ ($S'$ \emph{accepts} $A_1$) 
iff $A_2\in\semcai{S'}{\E'}$  ($S'$ \emph{accepts} $A_2$). 
When $S'=x$, if $i=0$ then both arrays are rejected (i.e., are not \emph{accepted}),
and when $i>0$ the result follows by induction on $i$.
Any conditional typed assertion (CTA) that is unrelated to arrays accepts both arrays,
while 
$\Uni$, and all $\Con$ and $\Enu$ in $D$, reject both of them, by construction.
Since they have the same length, $\CIte_j^l$ will not distinguish the two.
Consider any $S' = \Ite{S_1\itdots S_n}{S''}$ and assume that it accepts~$A_1$.
This means that each $S_i$ accepts~$\J_0$ and that $S''$ accepts both $\J_0$ and $\J_1$,
hence $A_2\in\semcai{S'}{\E'}$ as well. In the same way we prove that 
$A_2\in\semcai{S'}{\E'} \Implies A_1\in\semcai{S'}{\E'}$.
If $S'=\Cont{S''}$, then it cannot distinguish the two arrays since they contain the same elements.
If $S'=S_1 \And S_2$, or $S'=S_1 \Or S_2$, or $S'=\Not S_1$, we know, by induction, that~$S_1$ is not able to 
distinguish $A_1$ from $A_2$ and the same holds for~$S_2$, hence no boolean combination of $S_1$ and $S_2$
may distinguish $A_1$ from $A_2$.
To sum up, for every subterm $S'$ of $D$ we have that 
$\semcai{S'}{\E'}$ does not distinguish the two arrays, and hence the limit $\semca{S'}{\E'}$ does not distinguish them; therefore, 
$\semca{D}{}$ cannot be equivalent to $\Ex_2^{\Inf} S$.
The case $l > 2$ and the case $j\neq\Inf$ are proved in the same way, adjusting the
number of copies of $\J_1$ in the tails of $A_1$ and $A_2$.
\end{appendixproof}


\subsection{$\Ex_i^j S$: negation closure of array operators}
We have seen that negation closure for $\Ite{S_1\itdots S_n}{S_a}$ requires the dual operators
$\CIte_i^{\Inf}$ and  $\ContAfter{n}{S}$, 
but the latter cannot be expressed in the algebra without negation
(Theorem \ref{the:notitems}).
When we enrich the algebra with the $\Ex_i^j S$ operator, 
the situation changes completely. First of all, $\Ex_i^j S$ immediately subsumes the two  
operators $\Cont{S}$ and $\CIte_i^j$:
$$\begin{array}{lllllll}
\Cont{S} & = & \Ex_1^{\Inf} S 
\qquad\qquad\qquad
\CIte_i^j & = & \Ex_i^j \True 
\end{array}
$$

\oldversion{
More interestingly, $\Ex_i^j S$ also allows one to encode the $\ContAfter{n}{S}$ operator, as follows.
We define a function \SIf(Bit,$S$)  such that
\SIf(1,$S$) = $S$ and 
\SIf(0,$S$) = $\Not S$.
The operator $\ContAfter{n}{S}$ can be defined by the following disjunction with $2^{n}$ cases, one for each 
different bitmap of length $n$. 
Intuitively, for each such bitmap $bm$, the \IteK\ assertion states that the head of the array contains 
$\Sum(bm)$ elements that satisfy $S$ (in the positions $i$ where $bm[i]=1$), 
and $\Ex$ requires the global presence of at least $\Sum(bm)+1$ such elements.
The complete formulas says: $\J$ contains elements satisfying $S$ in the tail if there exists one bitmap $bm$
that specifies the positions of the $\Sum(\mathit{bm})$ elements in the head that satisfy $S$, and $\J$ contains,
globally, at least $\Sum(\mathit{bm})+1$ elements that satisfy $S$.
\[
\small
\begin{array}{lllllll}
\ContAfter{n}{S} = 
\quad\BigOr_{\mathit{bm} \in \BM}  \Ite{\SIf(bm[1],S)\itdots \SIf(bm[n],S)}{\True} \And  \Ex_{\Sum(\mathit{bm})+1}^{\Inf} S\ )
\end{array}
\]
}


More interestingly, $\Ex_i^j S$ also allows one to encode the $\ContAfter{n}{S}$ operator, as follows.
We define $I(l,p)$ as in Section \ref{sec:notelimination},
and we define 
a set of variables $U_{l,p,u}$, 
such that:
$$[\J_1,\ldots,\J_m]\in\semca{U_{l,p,u}}{\E}
\ \Iff\ 
| \setst{j}{j\in I_n(l,p) \And j\leq n \And \J_j \in \semca{S}{\E}}| \leq u$$
i.e., $\J\in\semca{U_{l,p,u}}{\E}$ implies that the number of elements $\J_j$ of $\J$ whose position
is in $I(l,p)$ but not in the tail $\Set{n+1,\ldots,2^{\LN}}$, and such that $\J_j\in \semca{S}{\E}$, is less than $u$.

In the first two lines, we deal with halving intervals of length $2^0=1$.
The second line ensures that all positions greater than $n+1$ will be ignored.
The third lines splits a generic halving interval $I$ in two halves, $I_1$ and $I_2$, 
and uses the same technique as in Section \ref{sec:notelimination}
to express $U_{l,p,u}$ in terms of $U_{l-1,2p-1,i}$ and $U_{l-1,2p,u-i}$.
The last line says that one $\J\in \semca{S}{\E}$ is contained in the array after position
$n$ iff, for some $i$, at most $i$ elements with $\J\in \semca{S}{\E}$ are found in positions $\SetTo{n}$, and the
array contains at least $i+1$ elements  with $\J\in \semca{S}{\E}$.
$$
\begin{array}{lllllll}
\multicolumn{3}{l}{
\Ee(S,n) =
} \\
( U_{0,p,0} &:& \IteAt{p}{\Not S}  &  
   1 \leq p \leq n    \\
\ U_{0,p,0} &:& \True  &  
   n+1 \leq p \leq 2^{\LN}    \\
\ U_{l,p,u} 
    &:&\BigOr_{0\leq i \leq u} ( U_{l-1,2p-1,i} \And  U_{l-1,2p,u-i} ) &
    1\leq l \leq q,\ 1 \leq p \leq 2^{\LN-l}, \\
    & & &   0\leq u \leq 2^{l}-1 \\
    )\\
\multicolumn{4}{l}{
( \ \ContAfter{n}{S}\ ,\ \E\ )\ \  = 
\ \  (\ \BigOr_{0 \leq i \leq n} ( U_{\LN,1,i} \And \Ex_{i+1}^{\Inf} S )\ ,\ \E \cup \Ee(S,n)\ )
} 
\end{array}
$$

Hence, $\Ex_i^j S$ is expressive enough to express $\ContAfter{n}{S}$, and thus to express negation
of $\IteK$, although at the cost of a complex encoding (of size $O(n^2)$).

Finally, we observe that $\Ex_i^j S$ is self-dual, so that, while it solves the problem of not-elimination for the
$\IteK$ operator,
it does not introduce any new not-elimination issue. The self-duality of $\Ex_i^j S$ is expressed by the following equation,
where $\Ex_{0}^{i-1}S$ is just $\False$ when $i=0$, and $\Ex_{j+1}^{\Inf}S$ is just $\False$ when $j=\Inf$.
$$
\Not(\Ex_{i}^{j}S) \ =\ 
\Type(\Arr) \And (\Ex_{0}^{i-1}S \Or \Ex_{j+1}^{\Inf}S) 
$$

While $\ContAfter{n}{S}$ is strictly less expressive than $\Ex_i^j S$, it seems to be more compact,
in the sense that we could not find any way to express $\ContAfter{n}{S}$ using $\Ex_i^j S$
with a linear-size expression.

Thus, while {\jsonsch} {\VerSix} needs four new operators --- $\PReq$, $\NotMof$, $\NotUni$,
and $\ContAfterK$ --- in order to become closed under negation, after $\Ex_{j}^{l}S$ is added, we only need three of them.


%% file: experiments.tex

We have implemented the algebra (with
the $\Ex_i^j S$ operator, to express negation of $\IteK$),
as well as the not-elimination algorithm. 
An interactive tool has been presented as a demo~\cite{DBLP:conf/edbt/AttoucheBCFGLSS21}, and is accessible online.\footnote{\url{https://jsonschematool.ew.r.appspot.com/}}
Our Java implementation comprises about $110K$ lines of code
 and uses the Brics automata library~\cite{bricks_automaton}
 for handling regular expressions. 
We performed two experiments, as described next.

\paragraph{Validation benchmark.} 
To assess the correctness of our implementation, we rely on schemas from the
JSON Schema Test Suite\footnote{\url{https://github.com/json-schema-org/JSON-Schema-Test-Suite}}  (commit hash \#8daea3f4).
This test suite is the de-facto standard benchmark for JSON Schema validators and
comprises 216 hand-crafted schemas for~\VerSix.
Each schema is encoded in just a few lines of code and targets a specific operator.
We have excluded all schemas with  advanced constructs that 
our implementation does not yet support (such as \verb!format!),
or with string escapings that our prototype cannot yet robustly parse,
but which our implementation could easily be extended to support.
We further removed schemas with references to external files,
which we cannot resolve. 
We then performed negation-elimination on the remaining 185 schemas,
having injected negation above the root.
Our implementation correctly eliminates negation. 

\paragraph{Real-world schemas.}
Our second experiment applies our algorithm to real-world schemas.
We have crawled GitHub for open source {\jsonsch} documents,
and retrieved over 80K files. As can be expected,
we encountered a multitude of problems in processing these non-curated,
raw files: files with syntactic errors,
files which do not comply to any JSON Schema draft,
and files with external references that we are unable to resolve.
We encountered troublesome string encodings, 
as well as patterns with forward/backward references
that we cannot represent with our automaton library.
Further, we encountered user-defined keywords that our implementation
cannot yet handle.

Notably, there is a large share of duplicate schemas,
with small variations in syntax and semantics, but evidently
versions of the same schema.
We rigorously removed such files, 
eliminating schemas with the same occurrences of keywords,
condensing the corpus down to 15\%.
Within the final collection of 2,229 files,
we are confident that they indeed represent \emph{individual} schemas
which vary in size from a few KBytes up to 0.4~MBytes, with an average of approx.~173 lines 
of code after pretty-printing.
Again, we inject negation above the document root 
and perform negation-elimination.

We ran our experiments on a PC with a i7-6700, 3.40GHz  CPU, 8~cores, 32~GB of main memory, 256 GB SDD. 
The tables below report the average runtime in milliseconds
 and the average runtime per KB (to account for different file sizes).
We further report the size ratio~(SR), 
i.e., the number of characters required for encoding the algebraic representation of the output schema,
versus the number of characters for the algebraic representation of the input schema. 
This is an indicator of the size increase due to not-elimination.

\begin{center}
\small 
\begin{tabular}{cc}

\begin{tabular}{ r  r  }
\toprule
\multicolumn{2}{c}{runtime (ms)} 
\\
 avg  &  avg/KB
 \\
\midrule
 2.77 & 0.46 
 \\
\bottomrule
\end{tabular}

&

\begin{tabular}{rr}
\toprule 
\multicolumn{2}{c}{size ratio}
\\
 avg & max 
 \\
\midrule
 2.78 & 27
 \\
\bottomrule
\end{tabular}

\end{tabular}

\end{center}

\code{
select coalesce(key,'*'), count(*)
from edftree
where path similar to '
group by rollup(key);

select count(distinct line)
from edftree
where path similar to '

wc -l files/* | sort -n 
}

\paragraph{Discussion.}
Translation to the algebra, combined with not-elimi\-nation, 
is in the sub-second range, which we consider acceptable. 
%
On average, not-elimination increases the size 
by a factor below~3. 
The maximum size ratio  is caused by not-elimination over enumerations with over 200 items.
Nevertheless, we observe linear growth.

Our experiments show that not-elimination
is indeed feasible on real-world {\jsonsch} documents.
While our prototype cannot yet handle all specific language constructs,
the current limitations are merely technical.
One unique selling point is that our approach fully supports negation and recursion
(even in combination), which is often a conceptual limitation of algorithms and tools designed
for {\jsonsch} processing (e.g., \cite{FruthJSCStudy}).

%% file: relatedwork.tex

In an empirical study~\cite{usage_of_not,DBLP:journals/corr/abs-2107-08677} over thousands of real-world schemas~\cite{schema_corpus}, we have analyzed usage patterns of the negation operator.
While we find occurrences of \verb!not! to be rare, we have found usage patterns of this operator to be subtle,
and often difficult to understand.

The problem of negation closure of {\jsonsch}, that is, the precise study of the duality among couples of
structural operators, does not seem to have been studied before.

Habib et al.~\cite{DBLP:conf/issta/HabibSHP21} study  schema inclusion for {\jsonsch}.  
Their algorithm is based on a form of not-elimination, hence showing how useful this technique is in practice
(notably, even for schemas that do not use the negation operator to start with).
They introduce many interesting techniques, but they only implement a limited form of not-elimination, since they
do not extend {\jsonsch} operators. Specifically, they do not address recursive definitions, 
although these are rather extensively used in real-world schemas~\cite{10.1007/978-3-030-34146-6}.


Indeed, our not-elimination algorithm for {\jsonsch} is the first to deal with the combination of negation and
recursive variables~\cite{FruthJSCStudy}, where we use a \emph{completion} technique that we believe to be original. 
The combination of negation and recursion has been deeply studied in the context of logic
languages, but these results cannot be easily transferred to {\jsonsch}, because of the different nature 
of these languages. For example, languages in the Prolog/Datalog family describe relations, while {\jsonsch} describes
sets. Moreover, variables in relational languages denote elements, while in {\jsonsch} a variable denotes a set,
like in Monadic Second Order logic (MSO). However, in MSO, variables are subject to quantification, 
while here, variables are only used to express recursion.
A logic language where variables denote sets, and are used for recursion rather than for quantification, 
is the $\mu$-calculus \cite{DBLP:journals/tcs/Kozen83}, which has been used to interpret 
{\jsonsch} in \cite{DBLP:conf/pods/BourhisRSV17}.
However, classical $\mu$-calculus techniques cannot be immediately transferred to this context,  since
$\mu$-calculus does not allow the presence of recursive variables below an odd number of negations, but they are
allowed by the {\jsonsch} standard, if recursion is guarded. 

\hide{The first effort to formalize the semantics of {\jsonsch} was by Pezoa et al.\  \cite{DBLP:conf/www/PezoaRSUV16}, whose goal was to lay the foundations of the {\jsonsch}  proposal by studying its expressive power and the complexity of the validation problem. 

Along the lines of this work, Bourhis et al.\  \cite{DBLP:conf/pods/BourhisRSV17} introduced the modal logic JSL  capturing the fourth draft of JSON Schema to investigate the complexity of the evaluation and the satisfiability problem. 
They use a technique of finite unfolding in order to interpret schemas
with negative recursion, and they prove that their formalization is as  expressive  as the JSON Schema language by describing the translation from JSON Schema to their logic, and back.
}

Works by  Pezoa et al.\  \cite{DBLP:conf/www/PezoaRSUV16} and Bourhis et al.\  \cite{DBLP:conf/pods/BourhisRSV17} have already been commented in the introduction. The semantics that we provide is not that different from that given in  \cite{DBLP:conf/pods/BourhisRSV17}, by means of the JSL modal logic. The main difference is that in  \cite{DBLP:conf/pods/BourhisRSV17} authors  translate {\jsonsch} into a formalism very far from it, 
while we directly deal with {\jsonsch} itself, although we provided an algebraic syntax, since we are interested in building practical tools that manipulate {\jsonsch} at the source level.